\documentclass[12pt,epsf]{article}

\usepackage{times}
\usepackage{graphicx}
\usepackage{amsfonts}
\usepackage{amsmath}
\usepackage{amssymb}
\usepackage{amstext}
\usepackage{amsthm}

\usepackage{epsfig}
\usepackage{color}
\psfigdriver{dvips}
\usepackage{latexsym}
\usepackage[matrix,curve]{xypic}
\usepackage{rotating}
\rotdriver{dvips}

\begin{document}

\baselineskip 6mm
\renewcommand{\thefootnote}{\fnsymbol{footnote}}

%------------ Hyun Seok's macro's, etc  -----------

\newcommand{\nc}{\newcommand}
\newcommand{\rnc}{\renewcommand}

%\headheight=0truein
%\headsep=0truein
%\topmargin=0truein
%\oddsidemargin=0truein
%\evensidemargin=0truein
%\textheight=9truein
%\textwidth=6.5truein

\rnc{\baselinestretch}{1.24}    % 1.5 spacing btwn text lines
\setlength{\jot}{6pt}       % spacing btwn the rows of an eqnarray
\rnc{\arraystretch}{1.24}   % spacing btwn the rows of a non-eqn array

%%%%%%%%%%%%%%%%%%%%%% Equation Numbering %%%%%%%%%%%%%%%%%%%%%%%
\makeatletter
\rnc{\theequation}{\thesection.\arabic{equation}}
\@addtoreset{equation}{section}
\makeatother

%%%%%%%%%%%%%%%%%%%%%%%%%%%%%%%%%%%%%%%%%%%%%%%%%%%%%%%%%%%%%%%%%
%                                                               %
%                NEW COMMANDS AND MACROS                        %
%                                                               %
%%%%%%%%%%%%%%%%%%%%%%%%%%%%%%%%%%%%%%%%%%%%%%%%%%%%%%%%%%%%%%%%%

%%%%% Simplify some frequently used LaTeX commands %%%%%

\nc{\be}{\begin{equation}}

\nc{\ee}{\end{equation}}

\nc{\bea}{\begin{eqnarray}}

\nc{\eea}{\end{eqnarray}}

\nc{\xx}{\nonumber\\}

\nc{\ct}{\cite}

\nc{\la}{\label}

\nc{\eq}[1]{(\ref{#1})}

\nc{\newcaption}[1]{\centerline{\parbox{6in}{\caption{#1}}}}

\nc{\fig}[3]{

\begin{figure}
\centerline{\epsfxsize=#1\epsfbox{#2.eps}}
\newcaption{#3. \label{#2}}
\end{figure}
}

%%% Caligraphic letters %%%%

\def\CA{{\cal A}}
\def\CC{{\cal C}}
\def\CD{{\cal D}}
\def\CE{{\cal E}}
\def\CF{{\cal F}}
\def\CG{{\cal G}}
\def\CH{{\cal H}}
\def\CK{{\cal K}}
\def\CL{{\cal L}}
\def\CM{{\cal M}}
\def\CN{{\cal N}}
\def\CO{{\cal O}}
\def\CP{{\cal P}}
\def\CR{{\cal R}}
\def\CS{{\cal S}}
\def\CU{{\cal U}}
\def\CV{{\cal V}}
\def\CW{{\cal W}}
\def\CY{{\cal Y}}
\def\CZ{{\cal Z}}

%%% Double line letters %%%

\def\IB{{\hbox{{\rm I}\kern-.2em\hbox{\rm B}}}}
\def\IC{\,\,{\hbox{{\rm I}\kern-.50em\hbox{\bf C}}}}
\def\ID{{\hbox{{\rm I}\kern-.2em\hbox{\rm D}}}}
\def\IF{{\hbox{{\rm I}\kern-.2em\hbox{\rm F}}}}
\def\IH{{\hbox{{\rm I}\kern-.2em\hbox{\rm H}}}}
\def\IN{{\hbox{{\rm I}\kern-.2em\hbox{\rm N}}}}
\def\IP{{\hbox{{\rm I}\kern-.2em\hbox{\rm P}}}}
\def\IR{{\hbox{{\rm I}\kern-.2em\hbox{\rm R}}}}
\def\IZ{{\hbox{{\rm Z}\kern-.4em\hbox{\rm Z}}}}

%%% Greek letters %%%

\def\a{\alpha}
\def\b{\beta}
\def\d{\delta}
\def\ep{\epsilon}
\def\ga{\gamma}
\def\k{\kappa}
\def\l{\lambda}
\def\s{\sigma}
\def\t{\theta}
\def\w{\omega}
\def\G{\Gamma}

%%%%% Mathematical Symbols

\def\half{\frac{1}{2}}
\def\dint#1#2{\int\limits_{#1}^{#2}}
\def\goto{\rightarrow}
\def\para{\parallel}
\def\brac#1{\langle #1 \rangle}
\def\curl{\nabla\times}
\def\div{\nabla\cdot}
\def\p{\partial}

%%%%% Roman pont in math

\def\Tr{{\rm Tr}\,}
\def\det{{\rm det}}

%%%%% Special Letters

\def\vare{\varepsilon}
\def\zbar{\bar{z}}
\def\wbar{\bar{w}}
\def\what#1{\widehat{#1}}

%%%%% For this paper only

\def\ad{\dot{a}}
\def\bd{\dot{b}}
\def\cd{\dot{c}}
\def\dd{\dot{d}}
\def\so{SO(4)}
\def\bfr{{\bf R}}
\def\bfc{{\bf C}}
\def\bfz{{\bf Z}}

\begin{titlepage}

%---------------- preprint number ---------------

\hfill\parbox{3.7cm} {{\tt arXiv:1101.1357}}

\vspace{15mm}

\begin{center}
%------------------------ title ------------------------
{\Large \bf  Yang-Mills Instantons from Gravitational Instantons}

\vspace{10mm}
%---------------- authors and addresses ----------------

John J. Oh${}^a$\footnote{johnoh@nims.re.kr}, Chanyong Park${}^b$\footnote{cyong21@sogang.ac.kr}
and Hyun Seok Yang${}^c$\footnote{hsyang@ewha.ac.kr}
\\[10mm]

${}^a$ {\sl National Institute for Mathematical Sciences, Daejeon 305-390, Korea}

${}^b$ {\sl Center for Quantum Space-Time, Sogang University, Seoul 121-742, Korea}

${}^c$ {\sl Institute for the Early Universe, Ewha Womans University, Seoul 120-750, Korea}

\end{center}

\thispagestyle{empty}

\vskip1cm

%----------------------- abstract ----------------------

\centerline{\bf ABSTRACT}
\vskip 4mm
\noindent

We show that every gravitational instantons are $SU(2)$ Yang-Mills
instantons on a Ricci-flat four manifold although the reverse is not necessarily true.
It is shown that gravitational instantons satisfy exactly the same self-duality equation
of $SU(2)$ Yang-Mills instantons on the Ricci-flat manifold
determined by the gravitational instantons themselves.
We explicitly check the correspondence with several examples and
discuss their topological properties.
\\



\vspace{1cm}

\today

\end{titlepage}

\renewcommand{\thefootnote}{\arabic{footnote}}
\setcounter{footnote}{0}

\section{Introduction}

An instanton in gauge theories is a topologically nontrivial solution described by
a self-dual or anti-self-dual connection with a finite action.
Such instantons play an important role in the nonperturbative dynamics of gauge theories,
in particular, to understand the vacuum structure of quantum field theories \cite{rajaraman}. One of the most powerful uses of instantons in
recent years is in the analysis of strongly coupled gauge dynamics where they play a
key role in unraveling the plexus of entangled dualities that relates different theories.
One of the highlights is the remarkable theory of Seiberg and Witten \ct{swn=2} which determines the low-energy behavior of $\mathcal{N}$ = 2 supersymmetric gauge theories exactly. In $\mathcal{N}$ = 2 supersymmetric gauge theories, the instantons lead
to quantum corrections for the metric on the moduli space of vacua.

A semi-classical evaluation of the path integral requires us to find the complete set of finite-action configurations which minimize the Euclidean action.
In pure Yang-Mills theory, the complete set of self-dual gauge fields of arbitrary
topological charge $k$ can be obtained by solving some
quadratic matrix equations, known as the Atiyah-Drinfeld-Hitchin-Manin (ADHM)
equations \ct{adhm}, which are a set of nonlinear algebraic equations constraining a matrix
of moduli parameters. It can be shown \cite{inst-review1} that the functional integral in the semi-classical approximation reduces to an integral over the instanton moduli space in each instanton sector. In principle the low-energy effective action can also be calculated
from first principles via conventional semi-classical methods using instantons.

It is known that the instanton calculus in supersymmetric theories is fully controllable
when the theories are weakly coupled. This leads to the idea of testing
the Seiberg-Witten theory by calculating the instanton effects and
comparing these expressions with those extracted from the Seiberg-Witten curve.
For reviews, see, for example, \cite{inst-review1,inst-review2}. Since the integral over a generic instanton moduli space is too complicated to be done directly,
it was fully accomplished only recently by using the localization technique
and considering the resolution of the instanton moduli space via the ADHM construction relevant to a noncommutative gauge theory \cite{nekrasov}.
It has been checked \cite{nek-oko} that the results computed using the method
of localization perfectly agree with the Seiberg-Witten solution
for $\mathcal{N}$ = 2 supersymmetric gauge theories.

On the mathematical side, instantons lie at the heart of the recent works on the topology
of four-manifolds \cite{donaldson}.
In particular, Donaldson used the moduli space of instantons over a
differentiable four-manifold to construct topological invariants of the four-manifold
and showed that the moduli spaces of instantons often carry nontrivial and
surprising information about the background manifold.

One would like to extend the path integral approach to include gravitation.
Although the Euclidean gravitational action is not positive-definite even for real
positive-definite metrics, one can evaluate the functional integral by first looking for
non-singular stationary points of the action functional and expand about them. Such critical points are finite action solutions to the classical field equations called ``gravitational instantons," the gravitational analogue of Yang-Mills instantons \cite{gi-hawking}. These are defined as complete, non-singular, and positive-definite metrics which are self-dual or anti-self-dual metrics of vacuum Einstein equations \cite{egh-report}.
One can show \cite{gibbs-pope} that the self-dual or anti-self-dual metrics are local minima of the action among metrics with zero scalar curvature.

In general relativity, the Lorentz group appears as the structure group acting on
orthonormal frames in the tangent space of a Riemannian manifold $M$ \cite{big-book}.
Under a local Lorentz transformation which is the orthogonal rotation group $O(4)$,
a matrix-valued spin connection ${\omega^A}_B = {{\omega_M}^A}_B dx^M$
plays a role of gauge fields in $O(4)$ gauge theory.
From the $O(4)$ gauge theory point of view, the Riemann curvature tensors precisely
correspond to the field strengths of the $O(4)$ gauge fields ${{\omega_M}^A}_B$.
(More details will be explained in Sect. 3.) Since the group $O(4)$ is a direct product of
normal subgroups $SU(2)_L$ and $SU(2)_R$, i.e. $O(4)= SU(2)_L \times SU(2)_R$,
the four-dimensional Euclidean gravity, when formulated as the $O(4)$ gauge theory, will
basically be two copies of $SU(2)$ gauge theories.

As we summarized above, Yang-Mills instantons are important to determine the vacuum structure of quantum field theories and the ADHM construction provides a description of all instantons on
$\mathbb{R}^4$ in terms of algebraic data. One would expect that gravitational instantons play a similar substantial role in quantum gravity although the quantum aspect of general relativity has encountered long-standing difficulties because there is hardly any common ground between general relativity and quantum mechanics. The well-known divergences
in a quantum theory of gravity suggest that a field theory of gravity like as Einstein's
general relativity is a purely low-energy or large-distance approximation to
some more fundamental theory. Therefore, the gauge theory formulation of gravity
may be helpful to glimpse some basic structures of such a fundamental theory because
nonperturbative and quantum aspects about gauge theories are relatively well-known.

Whereas gravity is different from gauge theory in several marked ways,
underlying mathematical structures are very similar to each other in many ways \cite{egh-report}. See, for example, the Table 1 in \ct{eh-ap}.
It was shown \cite{hitchin,kronheimer,gibbons,hitchin-cherkis} that certain classes of gravitational instantons such as the asymptotically locally Euclidean (ALE) and the asymptotically locally flat (ALF) hyper-K\"ahler four-manifolds can be constructed as a hyper-K\"ahler quotient of a finite-dimensional Euclidean space.\footnote{A general construction of essentially all known deformation classes of gravitational instantons was recently reported in \cite{kummer-gi}.} This construction is actually akin to the ADHM construction of Yang-Mills instantons on $\mathbb{R}^4$  \cite{adhm} and has a natural interpretation in terms of D-branes in string theory. Moreover, the hyper-K\"ahler quotient construction of Yang-Mills instantons on an ALE or ALF space \cite{kron-naka,italy,cherkis-alf,witten-alf} is a natural generalization of the original ADHM construction of instantons on flat space. The study of Yang-Mills theories on a curved manifold has recently received renewed attention because they are involved with effective field theories of D-brane and NS5-brane configurations \ct{sw-strong,ale-instanton}.

Now our motivation of this paper has surfaced. In this paper and its sequels,
we wish to go beyond a mere formal analogy between gravity and gauge theory
and try to answer to the following questions:

A. What is the precise relation between gravity and gauge theory variables?

B. How much are they parallel?

C. How is the topology of a Riemannian manifold $M$ encoded into gauge fields?

D. What are crucial differences?

E. Can it be applied to examine a quantum nature of gravity?

The paper is organized as follows. In Section 2, we will summarize Yang-Mills
instantons on a curved four-manifold to set our notation and
explain why Yang-Mills instantons on a Ricci-flat manifold is a solution of
the coupled equations in Einstein-Yang-Mills theory.

In Section 3, we will employ the decomposition in \cite{hsy-jhep} to explicitly realize
that the Lorentz group $O(4)$ is a direct product of normal subgroups $SU(2)_L$ and $SU(2)_R$, i.e. $O(4)= SU(2)_L \times SU(2)_R$.\footnote{See also
\ct{jjl-hsy} for geometric aspects of the decomposition according to the group
structure $O(4)= SU(2)_L \times SU(2)_R$.} It is then easy to show \cite{duff} that
the four-dimensional Euclidean gravity, when formulated as the $O(4)$ gauge theory,
will basically be two copies of $SU(2)$ gauge theories.
In particular, it can be shown that one of $SU(2)$'s decouples from the theory
when considering self-dual or anti-self-dual metrics called gravitational instantons.
As a result, one can show that gravitational instantons satisfy exactly
the same self-duality equation of $SU(2)$ Yang-Mills instantons
on the Ricci-flat manifold determined by the gravitational instantons themselves.
Therefore, every gravitational instantons can be interpreted
as self-gravitating $SU(2)$ Yang-Mills instantons
although the reverse is not necessarily true. This provides a powerful method
to find a particular class of Yang-Mills instantons on a general self-dual four manifold.

In Section 4, we will elucidate with explicit examples how it is always possible
to find Yang-Mills instantons on a Ricci-flat manifold $M$
using the prescription in Section 3 whenever a gravitational instanton solution is given.
Our method vividly realizes the Charap-Duff prescription \cite{duff}
for $SU(2)$ Yang-Mills instantons on a Ricci-flat manifold (see also \ct{het-inst}).
We will easily reproduce already known solutions
in literatures \ct{bcc1,bcc2,poyu,kim-yoon1} in this way
and also find new Yang-Mills instantons as a byproduct.

In Section 5, some issues about topological invariants for Riemannian manifolds
will be discussed. In the gravity side, there are two topological invariants \ct{egh-report} known as the Euler characteristic $\chi(M)$ and the Hirzebruch signature $\tau(M)$,
while, in the gauge theory side, there is a unique topological invariant up to a boundary term given by the Chern class of gauge bundle. The correspondence
between gravitational and Yang-Mills instantons then implies that the two topological
invariants for gravitational instantons should be related to each other.
We conjecture a possible relation between $\chi(M)$ and $\tau(M)$ by inspecting several known results in literatures \ct{gh-cmp,gpr-index,gibb-perry}.

In Section 6, we draw our conclusions and discuss open issues for future works.

Finally, we set up our index notation which is especially useful for the explicit
calculation in Section 4; otherwise diverse spaces we are considering would lead
to some confusions.

{\bf Index notation} We employ the following index convention throughout the paper:

$\bullet \; M, N, P, Q, \cdots = 1, \cdots, 4 \;$: world (curved space) indices,

$\bullet \; A, B, C, D, \cdots = \hat{1}, \cdots, \hat{4} \;$: frame (tangent space) indices,

$\bullet \; i, j, k, l, \cdots = 1, 2, 3 \;$: three-dimensional world indices,

$\bullet \; \hat{i}, \hat{j}, \hat{k}, \hat{l}, \cdots = \hat{1}, \hat{2}, \hat{3} \;$: three-dimensional frame indices,

$\bullet \; a, b, c, d, \cdots = \dot{1}, \dot{2}, \dot{3} \;$: $SU(2)$ Lie algebra indices.

\section{Yang-Mills Instantons on Riemannian Manifold}

Consider a curved four-manifold $M$ whose metric is given by
\be \la{back-space}
ds^2 = g_{MN}(x) dx^M dx^N.
\ee
Let $\pi: E \to M$ be an $SU(2)$ bundle over $M$ whose curvature is defined by
\bea \la{field-str}
F &=& dA + A \wedge A = \half F_{MN} (x) dx^M \wedge dx^N \xx
&=& \half \Big( \partial_M A_N - \partial_N A_M + [A_M, A_N] \Big) dx^M \wedge dx^N
\eea
where $A = A^a_M (x) T^a dx^M$ is a connection one-form of the vector bundle $E$.
The generators $T^a$ of $SU(2)$ Lie algebra satisfy the relation
\begin{equation}\label{su2-lie}
[T^a, T^b] = - 2 \varepsilon^{abc} T^c
\end{equation}
where we choose an unconventional normalization $\mathrm{Tr} T^a T^b = - 4 \delta^{ab}$
for later purpose.

Let us introduce at each spacetime point in $M$ a local frame of
reference in the form of four linearly independent vectors (vierbeins
or tetrads) $E_A = E_A^M \partial_M \in \Gamma(TM)$ which are chosen to be
orthonormal, i.e., $E_A \cdot E_B = \delta_{AB}$. The frame basis $\{ E_A \}$
defines a dual basis $E^A = E^A_M dx^M \in \Gamma(T^*M)$ by a natural pairing
\begin{equation} \label{dual-vector}
\langle E^A, E_B \rangle = \delta^A_B.
\end{equation}
The above pairing leads to the relation $E^A_M E_B^M = \delta^A_B$.
In terms of the non-coordinate (anholonomic) basis in $\Gamma(TM)$ or
$\Gamma(T^*M)$, the metric \eq{back-space} can be written as
\begin{eqnarray} \label{4-metric}
ds^2 &=& \delta_{AB} E^A \otimes E^B = \delta_{AB} E^A_M E^B_N
\; dx^M \otimes dx^N \nonumber
\\ &\equiv& g_{MN}(x) \; dx^M \otimes dx^N
\end{eqnarray}
or
\begin{eqnarray} \label{inverse-metric}
\Bigl(\frac{\partial}{\partial s}\Bigr)^2 &=& \delta^{AB} E_A \otimes E_B
= \delta^{AB} E_A^M E_B^N \; \partial_M \otimes \partial_N \nonumber
\\ &\equiv& g^{MN}(x)
\; \partial_M \otimes \partial_N.
\end{eqnarray}

Using the form language where $d = dx^M \partial_M = E^A E_A$
and $A = A_M dx^M = A_A E^A$, the field strength
(\ref{field-str}) of $SU(2)$ gauge fields in the non-coordinate basis takes the form
\begin{eqnarray} \label{form-field-noncod}
F &=& dA + A \wedge A = \frac{1}{2} F_{AB} E^A \wedge E^B \nonumber \\
&=&  \frac{1}{2} \Big( E_A A_B - E_B A_A + [A_A, A_B] + {f_{AB}}^C A_C \Big) E^A \wedge
E^B
\end{eqnarray}
where we used the structure equation
\begin{equation} \label{structure-eqn}
dE^A = \frac{1}{2} {f_{BC}}^A  E^B \wedge E^C.
\end{equation}
The frame basis $E_A = E_A^M \partial_M \in \Gamma(TM)$ satisfies the Lie algebra under the Lie bracket
\be \la{frame-lie}
[E_A, E_B] = - {f_{AB}}^C E_C
\ee
where
\begin{equation}\label{struct-f}
    f_{ABC} = E_A^M E_B^N (\partial_M E_{NC} - \partial_N E_{MC})
\end{equation}
are the structure functions in \eq{structure-eqn}.

Consider $SU(2)$ Yang-Mills theory defined on the Riemannnian manifold \eq{back-space}
whose action is given by
\begin{equation}\label{curved-ym}
    S_{YM} = - \frac{1}{16 g^2_{YM}} \int_{M} d^4 x \sqrt{g} g^{MP}g^{NQ}
    \mathrm{Tr} F_{MN} F_{PQ}.
\end{equation}
The self-duality equation for the action \eq{curved-ym} can be derived by observing
the following identity
\begin{equation} \label{curved-sdym}
    S_{YM} = - \frac{1}{32 g^2_{YM}} \int_M d^4 x \sqrt{g}  \mathrm{Tr} \Big(
    F_{MN} \mp \frac{1}{2}\frac{\varepsilon^{RSPQ}}{\sqrt{g}} g_{MR}g_{NS}  F_{PQ} \Big)^2 \mp
    \frac{1}{32 g^2_{YM}} \int_M d^4 x \varepsilon^{MNPQ} \mathrm{Tr} F_{MN} F_{PQ},
\end{equation}
where $\varepsilon^{MNPQ}$ is the metric independent Levi-Civita symbol with
$\varepsilon^{1234} = 1$.
Note that the second term in Eq. (\ref{curved-sdym}) is a topological term (total derivative)
and so does not affect the equations of motion.
Because the first term in Eq. (\ref{curved-sdym}) is positive-definite, the minimum of the action (\ref{curved-ym}) can be achieved by the self-dual gauge fields (instantons) satisfying
\begin{equation}\label{su2-instanton}
     F_{MN} = \pm \frac{1}{2}\frac{\varepsilon^{RSPQ}}{\sqrt{g}} g_{MR}g_{NS}  F_{PQ}.
\end{equation}
In the non-coordinate basis, the self-duality equation \eq{su2-instanton} can be written as the form
\begin{equation}\label{su2-frame}
     F_{AB} = \pm \frac{1}{2} {\varepsilon_{AB}}^{CD} F_{CD}
\end{equation}
with the field strength $F_{AB} = E_A^M E_B^N F_{MN}$ in \eq{form-field-noncod}.

It is easy to check that the $SU(2)$ instantons defined by \eq{su2-instanton}
automatically satisfy the equations of motion
\begin{equation}\label{eom-su2}
g^{MN} D_M F_{NP} = 0
\end{equation}
because we have the following relation from the self-duality \eq{su2-instanton}
\begin{equation}\label{self-eom}
 g^{MN} D_M F_{NP} = \mp \frac{1}{2} g_{PQ} \frac{\varepsilon^{QMNR}}{\sqrt{g}} D_M  F_{NR} =0
\end{equation}
where we used the Bianchi identity for the $SU(2)$ curvature \eq{field-str}, i.e.
\begin{equation}\label{jacobi-su2}
 \varepsilon^{MNPQ} D_N  F_{PQ} = 0.
\end{equation}
The covariant derivative in \eq{eom-su2} is with respect to both the Yang-Mills and
gravitational connections, i.e.
\begin{equation}\label{covder-gg}
 D_M F_{NP} = \partial_M F_{NP} - {\Gamma_{MN}}^Q F_{QP} - {\Gamma_{MP}}^Q F_{NQ} + [A_M, F_{NP}],
\end{equation}
where ${\Gamma_{MN}}^P$ is the Levi-Civita connection.

Now the problem we pose here is how to construct instanton solutions satisfying \eq{su2-instanton}. Several questions immediately arise.
Is it possible to find an instanton solution satisfying \eq{su2-instanton} on an arbitrary Riemannian manifold ? Or is there any constraint on the background manifold
for the existence of Yang-Mills instantons ? What is the moduli space of $SU(2)$ instantons
defined on a given four-manifold $M$ ?

We think the above questions are still open. Nevertheless, there are several examples on
Yang-Mills instantons defined on a curved four-manifold.
For example, the famous ADHM construction on $\mathbb{S}^4$ \ct{adhm},
Yang-Mills instantons on $\mathbb{C} P^2$ \ct{buchdahl}, $\mathbb{H} \times \mathbb{S}^2$
($\mathbb{H}=$ Poincar\'e half-plane) \ct{i-witten}, ALE \ct{kron-naka,italy} and
ALF spaces \ct{cherkis-alf,witten-alf}.
Also many other solutions have been constructed so far \ct{ete-hau,kim-yoon2,radu}.
See, for example, \ct{g-review,dunajski} for a review and references therein.
In particular, Taubes proved \ct{taubes} that all compact oriented four-manifolds admit
nontrivial instantons. But recently it was shown \ct{tsukamoto} that there exists a
noncompact four-manifold having no nontrivial instanton.
So far, we do not have a general description \`a la ADHM of all instantons satisfying
the self-duality \eq{su2-instanton}.

We will show that a large class of Yang-Mills instantons satisfying \eq{su2-instanton} or
\eq{su2-frame} can be solved by gravitational instantons. To be precise, we will show that
every gravitational instantons satisfy the self-duality equation \eq{su2-instanton}
for $SU(2)$ gauge fields on a Riemannian manifold defined by the gravitational
instanton itself. To prepare our setup, let us consider the case when $SU(2)$ Yang-Mills
and gravitational fields are both dynamically active. The total action is defined by
\begin{equation}\label{totol-action}
    S = S_{YM} + S_G
\end{equation}
where the Yang-Mills action $S_{YM}$ is given by \eq{curved-ym} and
the gravitational action is given by
\begin{equation}\label{einstein-hilbert}
    S_{G} = \frac{1}{16 \pi G} \int_{M} d^4 x \sqrt{g} R + {\rm surface \; terms}.
\end{equation}
The gravitational field equations read as
\begin{equation}\label{g-eom}
R_{MN} - \half g_{MN} R = 8 \pi G T_{MN}
\end{equation}
with
\begin{equation}\label{em-tensor}
T_{MN} =  \frac{1}{4 g^2_{YM}} \mathrm{Tr} \Big( g^{PQ}
     F_{MP} F_{NQ} - \frac{1}{4} g_{MN}  F_{PQ} F^{PQ} \Big).
\end{equation}

For an instanton solution satisfying Eq.\eq{su2-instanton},
the energy-momentum tensor \eq{em-tensor} identically vanishes, i.e.
$T_{MN} = 0$ and then Eq.\eq{g-eom} enforces the vacuum Einstein equations
\begin{equation}\label{vacuum-geom}
    R_{MN} = 0.
\end{equation}
Conversely, the reason that (anti-)self-dual Yang-Mills fields do not spoil Ricci-flatness
of a manifold is due to the vanishing of the Euclidean energy-momentum tensor \eq{em-tensor}.
Our interest is to solve the coupled equations \eq{su2-instanton} and
\eq{g-eom} simultaneously. Therefore, the four-manifold $M$ in
Eq.\eq{su2-instanton} should be Ricci-flat, i.e., satisfying the vacuum
Einstein equations \eq{vacuum-geom}.

\section{Gravitational Instantons}

Under local frame rotations in $O(4)$, the vectors transform
according to
\begin{eqnarray} \label{frame-rotation}
&& E_A(x) \to E_A^\prime(x) = E_B(x) {\Lambda^B}_A(x), \nonumber\\
&& E^A(x) \to {E^A}^\prime (x) =   {\Lambda^A}_B(x) E^B(x)
\end{eqnarray}
where ${\Lambda^A}_B(x) \in O(4)$. The spin connections
$\omega_M(x)$ then constitute gauge fields with respect to the local
$O(4)$ rotations
\begin{equation} \label{spin-so4}
\omega_M \to \Lambda \omega_M \Lambda^{-1} + \Lambda \partial_M \Lambda^{-1}
\end{equation}
and the covariant derivative is defined by
\begin{eqnarray} \label{spin-cov}
&& D_M E_A = \partial_M E_A - {{\omega_M}^B}_A E_B, \nonumber
\\ && D_M E^A = \partial_M E^A + {{\omega_M}^A}_B E^B.
\end{eqnarray}

The connection one-forms ${\omega^A}_B = {{\omega_M}^A}_B dx^M$
satisfy the Cartan's structure equations \cite{big-book},
\begin{eqnarray} \label{cartan-eq1}
T^A &=& dE^A + {\omega^A}_B \wedge E^B, \\
\label{cartan-eq2}
{R^A}_B &=& d{\omega^A}_B + {\omega^A}_C \wedge {\omega^C}_B,
\end{eqnarray}
where $T^A$ are the torsion two-forms and ${R^A}_B$ are the curvature two-forms.
In terms of local coordinates, they are given by
\begin{eqnarray} \label{cartan-torsion}
&& {T_{MN}}^A = \partial_M E_N^A - \partial_N E_M^A +
{{\omega_M}^A}_B E_N^B -
{{\omega_N}^A}_B E_M^B, \\
\label{cartan-curvature}
&& {{R_{MN}}^A}_B = \partial_M {{\omega_N}^A}_B - \partial_N
{{\omega_M}^A}_B+ {{\omega_M}^A}_C {{\omega_N}^C}_B -
{{\omega_N}^A}_C {{\omega_M}^C}_B.
\end{eqnarray}
Now we impose the torsion free
condition, ${T_{MN}}^A = D_M E_N^A - D_N E_M^A = 0$, to recover the
standard content of general relativity, which eliminates $\omega_M$
as an independent variable, i.e.,
\begin{eqnarray}\label{spin-conn}
\omega_{ABC} &=& E_A^M \omega_{MBC} =  \frac{1}{2} (f_{ABC} - f_{BCA} + f_{CAB}) \nonumber \\
&=& - \omega_{ACB}
\end{eqnarray}
where $f_{ABC}$ are the structure functions given by \eq{struct-f}. The spin connection
\eq{spin-conn} is related to the Levi-Civita connection as follows
\begin{equation}\label{levi-civita}
  {\Gamma_{MN}}^P = {{\omega_M}^A}_B E^P_A E^B_N + E^P_A \partial_M E_N^A.
\end{equation}

Since the spin connection $\omega_{MAB}$ and the curvature tensor
$R_{MNAB}$ are antisymmetric on the $AB$ index pair, one can
decompose them into a self-dual part and an anti-self-dual part as
follows \cite{hsy-jhep,jjl-hsy}
\begin{eqnarray} \label{spin-sd-asd}
&& \omega_{MAB} \equiv A_M^{(+)a} \eta^a_{AB} + A_M^{(-)a}
\bar{\eta}^a_{AB}, \\
\label{curvature-sd-asd}
&& R_{MNAB} \equiv F_{MN}^{(+)a} \eta^a_{AB} + F_{MN}^{(-)a}
\bar{\eta}^a_{AB},
\end{eqnarray}
where the $4 \times 4$ matrices $\eta^a_{AB}$ and ${\bar
\eta}^a_{AB}$ for $a=\dot{1},\dot{2},\dot{3}$ are 't Hooft symbols defined by
\begin{eqnarray} \label{tHooft-symbol}
&& {\bar \eta}^a_{\hat{i}\hat{j}} = {\eta}^a_{\hat{i}\hat{j}}
= {\varepsilon}_{a\hat{i}\hat{j}}, \qquad
\hat{i},\hat{j} \in \{\hat{1},\hat{2},\hat{3}\}, \nonumber\\
&& {\bar \eta}^a_{\hat{4}\hat{i}} = {\eta}^a_{\hat{i}\hat{4}}
= \delta_{a\hat{i}}.
\end{eqnarray}
Note that the 't Hooft matrices intertwine the group structure of
the index $a$ with the spacetime structure of the indices $A, B$. We
list some useful identities of the 't Hooft tensors \cite{hsy-jhep,jjl-hsy}
\begin{eqnarray} \label{self-eta}
&& \eta^{(\pm)a}_{AB} = \pm \frac{1}{2} {\varepsilon_{AB}}^{CD}
\eta^{(\pm)a}_{CD}, \\
\label{proj-eta}
&& \eta^{(\pm)a}_{AB}\eta^{(\pm)a}_{CD} =
\delta_{AC}\delta_{BD}
-\delta_{AD}\delta_{BC} \pm
\varepsilon_{ABCD}, \\
\label{self-eigen}
&& \varepsilon_{ABCD} \eta^{(\pm)a}_{DE} = \mp (
\delta_{EC} \eta^{(\pm)a}_{AB} + \delta_{EA} \eta^{(\pm)a}_{BC} -
\delta_{EB} \eta^{(\pm)a}_{AC} ), \\
\label{eta-etabar}
&& \eta^{(\pm)a}_{AB} \eta^{(\mp)b}_{AB}=0, \\
\label{eta^2}
&& \eta^{(\pm)a}_{AC}\eta^{(\pm)b}_{BC} =\delta^{ab}\delta_{AB} +
\varepsilon^{abc}\eta^{(\pm)c}_{AB}, \\
\label{eta-ex}
&& \eta^{(\pm)a}_{AC}\eta^{(\mp)b}_{BC} =
\eta^{(\mp)b}_{AC}\eta^{(\pm)a}_{BC}, \\
\label{eta-o4-algebra}
&& \varepsilon^{abc} \eta^{(\pm)b}_{AB} \eta^{(\pm)c}_{CD} =
    \delta_{AC} \eta^{(\pm)a}_{BD} - \delta_{AD} \eta^{(\pm)a}_{BC}
    - \delta_{BC} \eta^{(\pm)a}_{AD} + \delta_{BD} \eta^{(\pm)a}_{AC}
\end{eqnarray}
where $\eta^{(+)a}_{AB} \equiv \eta^a_{AB}$ and $\eta^{(-)a}_{AB} \equiv {\bar
\eta}^a_{AB}$.

Of course all these separations are due to the fact, $O(4)= SU(2)_L
\times SU(2)_R$, stating that any $O(4)$ rotations can be decomposed
into self-dual and anti-self-dual rotations. To be explicit,
for an infinitesimal $O(4)$ transformation,
i.e., ${\Lambda^A}_{B}(x) \approx {\delta^A}_{B} +  {\lambda^A}_{B}(x)$,
we can take the following decomposition
\begin{equation}\label{04-decomp}
    \lambda_{AB}(x) =  \lambda^a_{(+)} (x) \eta^a_{AB} + \lambda^a_{(-)}(x)
\bar{\eta}^a_{AB}
\end{equation}
where $\lambda^a_{(+)}(x)$ and $\lambda^a_{(-)}(x)$ are local gauge parameters
in $SU(2)_{L}$ and $SU(2)_{R}$, respectively.
To be specific, let us introduce two families of $4 \times 4$ matrices defined by
\begin{equation} \label{thooft-matrix}
[T^a_+]_{AB} \equiv \eta^a_{AB}, \qquad [T^a_-]_{AB} \equiv {\bar
\eta}^a_{AB}.
\end{equation}
According to the definition \eq{tHooft-symbol}, the matrix representation of the generators
in \eq{thooft-matrix} is given by
\begin{eqnarray}
&& T^{\dot{1}}_+ = \begin{pmatrix}
      0 & 0 & 0 & 1 \\
      0 & 0 & 1 & 0 \\
      0 & -1 & 0 & 0 \\
      -1 & 0 & 0 & 0 \\
       \end{pmatrix}, \quad
  T^{\dot{2}}_+ = \begin{pmatrix}
      0 & 0 & -1 & 0 \\
      0 & 0 & 0 & 1 \\
      1 & 0 & 0 & 0 \\
      0 & -1 & 0 & 0 \\
    \end{pmatrix}, \quad
  T^{\dot{3}}_+ = \begin{pmatrix}
      0 & 1 & 0 & 0 \\
      -1 & 0 & 0 & 0 \\
      0 & 0 & 0 & 1 \\
      0 & 0 & -1 & 0 \\
    \end{pmatrix}, \\
&& T^{\dot{1}}_- = \begin{pmatrix}
      0 & 0 & 0 & -1 \\
      0 & 0 & 1 & 0 \\
      0 & -1 & 0 & 0 \\
      1 & 0 & 0 & 0 \\
    \end{pmatrix}, \quad
  T^{\dot{2}}_- =  \begin{pmatrix}
      0 & 0 & -1 & 0 \\
      0 & 0 & 0 & -1 \\
      1 & 0 & 0 & 0 \\
      0 & 1 & 0 & 0 \\
    \end{pmatrix}, \quad
  T^{\dot{3}}_- =  \begin{pmatrix}
      0 & 1 & 0 & 0 \\
      -1 & 0 & 0 & 0 \\
      0 & 0 & 0 & -1 \\
      0 & 0 & 1 & 0 \\
    \end{pmatrix}.
\end{eqnarray}
Then Eqs. (\ref{eta^2}) and (\ref{eta-ex}) immediately show that
$T^a_\pm$ satisfy $SU(2)$ Lie algebras, i.e.,
\begin{equation} \label{thooft-su2}
[T^a_\pm, T^b_\pm] = - 2 \varepsilon^{abc} T^c_\pm,
\qquad [T^a_\pm, T^b_\mp] = 0.
\end{equation}
According to the definition \eq{thooft-matrix}, the self-duality \eq{self-eta} leads
to the important relation
\begin{equation}\label{self-dual-su2}
   [T^a_\pm]_{AB} = \pm \frac{1}{2} {\varepsilon_{AB}}^{CD} [T^a_\pm]_{CD}.
\end{equation}

The 't Hooft matrices in (\ref{thooft-matrix}) are two
independent spin $s = \frac{3}{2}$ representations of $SU(2)$ Lie
algebra. A deep geometrical meaning of the 't Hooft symbols is to
specify the triple $(I,J,K)$ of complex structures of $\mathbb{R}^4
\cong \mathbb{C}^2$ as the simplest hyper-K\"ahler manifold for a given orientation.
The triple complex structures $(I,J,K)$ form a quaternion which can
be identified with the $SU(2)$ generators $T^a_\pm$ in (\ref{thooft-matrix}) \ct{jjl-hsy}.

Now we introduce an $O(4)$-valued gauge field defined
by $A = A^{(+)a} T_+^a + A^{(-)a} T_-^a$ where $A^{(\pm)a} = A^{(\pm)a}_{M} dx^M \;
(a = 1, 2, 3)$ are connection one-forms on $M$ and
$T^a_\pm$ are Lie algebra generators of $SU(2)_L$ and $SU(2)_R$ satisfying \eq{thooft-su2}.
The identification we want to make is then given by
\begin{equation}\label{id}
    \omega = \frac{1}{2} \omega_{AB} J^{AB} \equiv A = A^{(+)a} T_+^a + A^{(-)a} T_-^a.
\end{equation}
Since the group $SO(4)$ is a direct product of normal subgroups
$SU(2)_L$ and $SU(2)_R$, i.e. $SO(4) = SU(2)_L \times SU(2)_R$,
we take the 4-dimensional defining representation of the Lorentz generators as follows
\begin{eqnarray}\label{rep-so4}
    [J^{AB}]_{CD} &=& \frac{1}{2} \Big(\eta^a_{AB} [T^a_+]_{CD} + \bar{\eta}^{a}_{AB} [T^{a}_-]_{CD} \Big) \xx
    &=& \frac{1}{2} \Big(\eta^a_{AB} \eta^a_{CD} + \bar{\eta}^{a}_{AB} \bar{\eta}^{a}_{CD} \Big),
\end{eqnarray}
where $T^a_+$ and $T^{a}_-$ are the $SU(2)_L$ and $SU(2)_R$ generators given by
Eq. \eq{thooft-matrix}. It is then easy to check using Eqs. \eq{thooft-su2} and \eq{eta-o4-algebra} or Eq. \eq{proj-eta} that the generators in Eq. \eq{rep-so4} satisfy
the Lorentz algebra. According to the identification \eq{id},
$SU(2)$ gauge fields can be defined from the spin connections
\begin{eqnarray}\label{spin-gauge}
    [\omega_M]_{CD} &=& \frac{1}{2}\omega_{MAB} [J^{AB}]_{CD} \xx
    &=& \Big( \frac{1}{2} \omega_{MAB} \eta^a_{AB} \Big) [T^a_+]_{CD}
    + \Big( \frac{1}{2} \omega_{MAB} \bar{\eta}^{a}_{AB} \Big)
    [T^{a}_-]_{CD} \xx
    &\equiv & A_M^{(+)a} [T^a_+]_{CD} + A_M^{(-)a}[T^{a}_-]_{CD}
    = [A_M]_{CD}.
\end{eqnarray}
That is, we get the decomposition \eq{spin-sd-asd} for spin connections.

Using the definition \eq{thooft-matrix}, the spin connection \eq{spin-sd-asd}
and the curvature tensor \eq{curvature-sd-asd} can be written as follows:
\begin{eqnarray} \la{dec-spin}
 \omega_{MAB} & = &  A^{(+)a}_M [T_+^a]_{AB} +  A^{(-)a}_M [T_-^a]_{AB}, \\
 \la{dec-curv}
  R_{MNAB} & = &  F^{(+)a}_{MN} [T_+^a]_{AB} +  F^{(-)a}_{MN} [T_-^a]_{AB},
\end{eqnarray}
where
\begin{equation} \label{gi-curvature}
F_{MN}^{(\pm)} = \partial_M A_N^{(\pm)} - \partial_N A_M^{(\pm)} +
[A_M^{(\pm)}, A_N^{(\pm)}].
\end{equation}
Using the Lie algebra (\ref{thooft-su2}), one can write the field strength
\eq{gi-curvature} as the component form
\begin{equation} \label{su2-curvature-com}
F_{MN}^{(\pm)a} = \partial_M A_N^{(\pm)a} - \partial_N A_M^{(\pm)a}
- 2 \varepsilon^{abc} A_M^{(\pm)b}A_N^{(\pm)c},
\end{equation}
which is precisely the same as Eq. \eq{field-str}.
Therefore, we see that $A_M^{(\pm)} = A_M^{(\pm)a}T^a_\pm$
can be identified with $SU(2)_{L,R}$ gauge fields and
$F_{MN}^{(\pm)} = F_{MN}^{(\pm)a}T^a_\pm$ with their field strengths.
Indeed one can also show that the local $O(4)$ rotations in (\ref{spin-so4})
can be represented as the gauge transformations of the $SU(2)$ gauge fields
$A_M^{(\pm)}$:
\begin{equation} \label{spin-su2}
A^{(\pm)}_M \to \Lambda_{(\pm)} A^{(\pm)}_M
\Lambda^{-1}_{(\pm)} + \Lambda_{(\pm)} \partial_M \Lambda^{-1}_{(\pm)}
\end{equation}
where $\Lambda_{(\pm)} (x) \equiv \exp(\lambda^a_{(\pm)}(x) T^a_\pm) \in SU(2)_{L,R}$
are group elements defined by Eq. \eq{04-decomp}.

Let us recall the symmetry property of curvature tensors
determined by the properties about the torsion and the tangent-space
group
\begin{equation} \label{curvature-anti-symm}
R_{ABCD} = - R_{ABDC} = -R_{BACD}
\end{equation}
where $R_{ABCD} = E_A^M E_B^N R_{MNCD}$. Also note that the
curvature tensors satisfy the first Bianchi identity
\begin{equation} \label{1st-bianchi}
R_{A[BCD]} \equiv R_{ABCD} + R_{ADBC} + R_{ACDB} = 0
\end{equation}
which is an integrability condition originated by the fact that the spin
connections (\ref{spin-conn}) are determined by potential
fields, i.e., vierbeins. It is easy to
see that the following symmetry can be derived by using
Eqs. (\ref{curvature-anti-symm}) and (\ref{1st-bianchi})
\begin{equation} \label{symm-two-riemann}
 R_{ABCD} = R_{CDAB}.
\end{equation}

The gravitational instantons are defined by the self-dual solution
to the Einstein equation
\begin{equation} \label{g-instanton}
R_{MNAB} = \pm \frac{1}{2} {\varepsilon_{AB}}^{CD} R_{MNCD}.
\end{equation}
Note that a metric satisfying the self-duality equation
(\ref{g-instanton}) is necessarily Ricci-flat because $R_{MN} \equiv {R_{MAN}}^A =
\pm \frac{1}{6} {\varepsilon_{N}}^{ABC} R_{M[ABC]} = 0$ and so automatically
satisfies the vacuum Einstein equations \eq{vacuum-geom}. Using the decomposition
(\ref{dec-curv}) and the relation \eq{self-dual-su2}, Eq.\eq{g-instanton} can be written as
\bea \la{half-flat}
F_{MN}^{(+)a} [T^a_+]_{AB} + F_{MN}^{(-)a}
[T^a_-]_{AB} &=& \pm \frac{1}{2} {\varepsilon_{AB}}^{CD} (F_{MN}^{(+)a} [T^a_+]_{CD} + F_{MN}^{(-)a} [T^a_-]_{CD}) \xx
&=&\pm (F_{MN}^{(+)a} [T^a_+]_{AB} - F_{MN}^{(-)a} [T^a_-]_{AB}).
\eea
Therefore we should have $F_{MN}^{(-)a} = 0$ for the self-dual case
with $+$ sign in Eq. \eq{g-instanton} while $F_{MN}^{(+)a} = 0$ for
the anti-self-dual case with $-$ sign and so imposing the
self-duality equation \eq{g-instanton} is equivalent to the
half-flat equation $F^{(\pm)a} = 0$.

A solution of the half-flat equation $F^{(\pm)} = 0$ is given by
$A^{(\pm)} = \Lambda_\pm d \Lambda^{-1}_\pm$ and then Eq.\eq{spin-su2} shows that
it is always possible to choose a self-dual gauge $A^{(\pm)a} = 0$.
Therefore, one can see the following important property.
If the spin connection is, for example, self-dual,
i.e. $A_M^{(-)}=0$, the curvature tensor is also self-dual, i.e.
$F_{MN}^{(-)}=0$. Conversely, if the curvature is self-dual, i.e.
$F_{MN}^{(-)}=0$, one can always choose a self-dual spin connection
by a suitable gauge choice since $F_{MN}^{(-)}=0$ requires that
$A_M^{(-)}$ is a pure gauge. In other words, in this self-dual gauge, the
problem of finding gravitational instantons is equivalent to one of
finding self-dual spin connections \cite{eh-ap}
\begin{equation} \label{sde-spin}
\omega_{MAB} = \pm \frac{1}{2}  {\varepsilon_{AB}}^{CD} \omega_{MCD}
\end{equation}
which is equivalent to the (anti-)self-dual gauge condition $A^{(\pm)a}_M = 0$
according to the decomposition \eq{dec-spin}. The gravitational
instantons defined by Eq.(\ref{g-instanton}) are then obtained by
solving the first-order differential equations defined by (\ref{sde-spin}).

The self-duality equations (\ref{g-instanton}) are imposed on the
second group indices $[CD]$ of the curvature tensor $R_{ABCD}$ and
they do not touch the first group indices $[AB]$. But note that the
first Bianchi identity (\ref{1st-bianchi}) reshuffles three indices
in $R_{ABCD}$ and the symmetry (\ref{symm-two-riemann}) is
consequently deduced. Thereby the self-duality condition for the
second group should necessarily be correlated to the one for the
first group \ct{duff}. In other words, because the Riemann curvature tensors
satisfy the symmetry property \eq{symm-two-riemann},
the gravitational instanton \eq{g-instanton} is equivalent to the self-duality
equation
\begin{equation} \label{gr-instanton}
R_{ABEF} = \pm \frac{1}{2} {\varepsilon_{AB}}^{CD} R_{CDEF}.
\end{equation}
Then, using the decomposition (\ref{dec-curv}) again, one
can similarly show that the gravitational instanton
(\ref{gr-instanton}) can be understood as an $SU(2)$ Yang-Mills
instanton defined by \eq{su2-frame}, i.e.
\begin{equation} \label{ym-instanton}
F^{(\pm)}_{AB} = \pm \frac{1}{2} {\varepsilon_{AB}}^{CD}
F^{(\pm)}_{CD}
\end{equation}
where $F^{(\pm)}_{AB} = F^{(\pm) a}_{AB} T^a_\pm = E_A^M E_B^N
F_{MN}^{(\pm)}$ are defined by Eq. (\ref{form-field-noncod}).
In a coordinate basis, the self-duality equation
(\ref{ym-instanton}) can be written as the form \eq{su2-instanton}
because one can deduce that
\begin{eqnarray} \label{ym-instanton-cod}
E_A^M E_B^N F^{(\pm)}_{MN} &=& \pm \frac{1}{2} {\varepsilon_{AB}}^{CD}
E_C^P E_D^Q F^{(\pm)}_{PQ} \nonumber \\
\Rightarrow F^{(\pm)}_{MN} &=& \pm \frac{1}{2} {\varepsilon_{AB}}^{CD}
E^A_M E^B_N E_C^P E_D^Q F^{(\pm)}_{PQ} \nonumber \\
&=& \pm \frac{1}{2}  g_{MR} g_{NS}\, \varepsilon^{ABCD}
E_A^R E_B^S E_C^P E_D^Q F^{(\pm)}_{PQ} \nonumber \\
&=& \pm  \frac{1}{2} \frac{\varepsilon^{RSPQ}}{\sqrt{g}} g_{MR} g_{NS}
 F^{(\pm)}_{PQ}
\end{eqnarray}
where $\sqrt{g} = \det E_M^A$.

Therefore, we see that gravitational instantons defined by Eq. \eq{g-instanton}
are solutions of both \eq{su2-instanton} and \eq{vacuum-geom} and so they
can be regarded as Yang-Mills instantons in the sense that the self-duality equation
of gravitational instantons can always be recast into exactly the same self-duality
equation as the $SU(2)$ Yang-Mills instantons on a Ricci-flat manifold.
But note that the Yang-Mills instantons as well as the four-dimensional metric
used to define Eq. \eq{ym-instanton-cod} are simultaneously determined
by gravitational instantons. Therefore, the self-duality in Eq. \eq{ym-instanton-cod}
cannot be interpreted as $SU(2)$ instantons in a fixed background.
Although every gravitational instantons satisfy the self-duality
equation \eq{su2-instanton} for Yang-Mills instantons on a Ricci-flat manifold,
the converse is not necessarily true: An $SU(2)$ instanton
on a Ricci-flat manifold is not always a gravitational instanton.
For example, Yang-Mills instantons on ALE spaces in \ct{kron-naka,italy}
and ALF spaces in \ct{cherkis-alf,witten-alf} consist of a more general class of
solutions than those obtained from ALE and ALF gravitational instantons.

As was pointed out above, the self-duality in Eq. \eq{ym-instanton-cod} should not be interpreted as $SU(2)$ instantons in a fixed background because we are solving the coupled
equations \eq{su2-instanton} and \eq{vacuum-geom}. We are not solving Eq. \eq{su2-instanton} on a non-dynamical background manifold. Note that the Yang-Mills action \eq{curved-ym}
is invariant under the conformal transformation
\begin{equation}\label{conf-tr}
    g_{MN} \mapsto \widetilde{g}_{MN} = \Omega^2(x) g_{MN},
\end{equation}
assuming that $F_{MN}$ are metric-independent. As a result, the self-duality equations \eq{su2-instanton} are also invariant under the transformation \eq{conf-tr}.
However the conformal transformation \eq{conf-tr} is no longer a symmetry of the coupled system
defined by the action \eq{totol-action} because the gravitational action \eq{einstein-hilbert}
is not invariant under the transformation \eq{conf-tr} and so breaks
the conformal symmetry. Furthermore the assumption that $F_{MN}$ are metric-independent
is no longer valid when gravity is coupled to Yang-Mills fields.
Therefore, Eq. \eq{ym-instanton-cod} does not have to be invariant
under the conformal transformation \eq{conf-tr}. Of course, this feature is consistent
with the fact that the Yang-Mills instantons satisfying Eq.\eq{ym-instanton-cod} are
defined by the Einstein-Yang-Mills action \eq{totol-action}.

We will finally check the claim that the gravitational instantons can be regarded as Yang-Mills instantons by showing that the former satisfies the same equations
as the latter. First, we show that the second Bianchi identity for curvature
tensors is reduced to the Bianchi identity for $SU(2)$ gauge fields:
\begin{equation} \label{2nd-bianchi}
\nabla_{[M} R_{NP]AB} = 0 \quad \Leftrightarrow  \quad D^{(\pm)}_{[M} F^{(\pm)}_{NP]} =
0,
\end{equation}
where the bracket $[MNP] \equiv MNP + NPM + PMN$ denotes the cyclic permutation of indices.
The covariant derivative on the left-hand side of Eq. \eq{2nd-bianchi} is defined by
\begin{equation}\label{cov-der-curvature}
\nabla_{M} R_{NPAB} = \partial_{M} R_{NPAB} - {\Gamma_{MN}}^{Q} R_{QPAB}
-  {\Gamma_{MP}}^{Q} R_{NQAB} - {{\omega_M}^C}_A R_{NPCB} - {{\omega_M}^C}_B R_{NPAC}
\end{equation}
and, on the right-hand side, it is given by Eq. \eq{covder-gg}.
Rewrite the covariant derivative \eq{cov-der-curvature} as the form
\begin{equation*}
\nabla_{M} R_{NPAB} = \partial_{M} R_{NPAB} - {\Gamma_{MN}}^{Q} R_{QPAB}
-  {\Gamma_{MP}}^{Q} R_{NQAB} + \omega_{MAC} R_{NPCB}
- R_{NPAC} \omega_{MCB}.
\end{equation*}
Using the decompositions (\ref{dec-spin}) and (\ref{dec-curv}) and
the commutation relations \eq{thooft-su2}, we get
\begin{eqnarray} \la{end-proof}
\nabla_{M} R_{NPAB}
&=& \Big(\partial_{M} F_{NP}^{(+)} - {\Gamma_{MN}}^{Q} F^{(+)}_{QP}
- {\Gamma_{MP}}^{Q} F^{(+)}_{NQ} + [A_{M}^{(+)}, F_{NP}^{(+)}] \Big)_{AB} \nonumber\\
&& + \Big(\partial_{M} F_{NP}^{(-)} - {\Gamma_{MN}}^{Q} F^{(-)}_{QP}
- {\Gamma_{MP}}^{Q} F^{(-)}_{NQ} + [A_{M}^{(-)}, F_{NP}^{(-)}] \Big)_{AB} \xx
&=& \Big( D^{(+)}_{M} F_{NP}^{(+)} + D^{(-)}_{M} F_{NP}^{(-)} \Big)_{AB}.
\end{eqnarray}
Therefore, we arrived at the result (\ref{2nd-bianchi}) that the
second Bianchi identity for curvature tensors is equivalent to the
Bianchi identity for $SU(2)$ Yang-Mills fields. Note that all the terms
containing the Levi-Civita connection in Eq.\eq{2nd-bianchi} are canceled each other.

After rewriting the self-duality equation \eq{gr-instanton} as
\begin{equation}\label{curv-g-instanton}
    R_{MNAB} = \pm \half \frac{\varepsilon^{RSPQ}}{\sqrt{g}} g_{MR} g_{NS} R_{PQAB},
\end{equation}
the covariant derivative is taken on both sides to yield
\begin{equation*}
g^{PM} \nabla_{P} R_{MNAB} = \mp \half \frac{{\varepsilon_N}^{RPQ}}{\sqrt{g}}
\nabla_R R_{PQAB} = 0,
\end{equation*}
where the Bianchi identity (\ref{2nd-bianchi}) was used. The relation \eq{end-proof} then
guarantees that the Yang-Mills equations
\begin{equation} \label{ym-eom}
g^{MN} D^{(\pm)}_M F^{(\pm)}_{NP} = 0
\end{equation}
will be satisfied accordingly.
So remarkably it turns out that gravitational instantons can actually be identified with Yang-Mills instantons in the sense that the gravitational and Yang-Mills instantons
satisfy mathematically the same self-duality equations.
But, as we discussed before, the self-duality equation (\ref{ym-instanton-cod})
must be interpreted as self-gravitating Yang-Mills instantons rather than
$SU(2)$ instantons on a rigid background.

\section{Yang-Mills Instantons from Gravitational Instantons}

We showed in the previous section that every gravitational instantons satisfy
the self-duality equation \eq{su2-instanton} on a Ricci-flat manifold
defined by the gravitational instanton itself.
We have constructed $SU(2)$ gauge fields as the projection of
the spin connection \eq{spin-so4} onto the self-dual part and the anti-self-dual part
by using the 't Hooft symbols. The embedding to relate gauge and spin connections
was suggested long ago by Charap and Duff \ct{duff}. (See also \cite{het-inst}.)
In this section, we will elucidate with explicit examples
how Yang-Mills instantons can be obtained from gravitational instantons.

To be specific, we want to find Yang-Mills instantons satisfying Eq. \eq{su2-instanton}
where the background metric $g_{MN}$ is a gravitational instanton
obeying Eq. \eq{gr-instanton}. First, we will calculate the spin connection \eq{spin-conn}
for a given gravitational instanton metric and then identify
$SU(2)$ gauge fields $A_M$ according to the identification \eq{id}.
As was shown in \eq{ym-instanton-cod}, the corresponding field strength $F_{MN}$ of
the $SU(2)$ gauge fields automatically satisfies the self-duality equation \eq{su2-instanton}
on a curved manifold $M$ whose metric is given by the gravitational instanton itself.

We will easily reproduce already known solutions
in literatures \ct{bcc1,bcc2,poyu,kim-yoon1} in this way.
As a byproduct, we will also find new Yang-Mills instantons on a curved manifold $M$.
It might be emphasized that it is always possible to find Yang-Mills instantons
on a Ricci-flat manifold $M$ by the same procedure whenever a gravitational instanton $M$
is given, as will be illustrated with several examples.
Here we refer to the index convention in Section 1.

\subsection{Gibbons-Hawking metric}

The Gibbons-Hawking metric \cite{gh-instanton} is a general class of
self-dual, Ricci-flat metrics with the triholomorphic $U(1)$
symmetry which describes a particular class (A-type) of ALE and ALF instantons.
The Gibbons-Hawking metric for gravitational multi-instantons
is given by
\begin{eqnarray} \la{gh-metric}
ds^2 &=& V^{-1}(x)(d\tau + q_{i}dx^{i})^2 + V(x)dx^{i}dx^{i} \xx
&\equiv& e^{2\psi} (d\tau + q_{i} dx^{i})^2 + e^{-2\psi} dx^{i}dx^{i},
\end{eqnarray}
where
\begin{equation} \label{harmonic-v}
V(x) =  e^{-2\psi(x)} = \epsilon + 2m\sum_{a=1}^{k}\frac{1}{|{x}^{i} - {x}_{a}^{i}|}
\end{equation}
with $\epsilon = 0$ for ALE instantons and $\epsilon = 1$ for ALF instantons.
Here we use the world index $M=(i,4=\tau)$ with $i=1,2,3$ and the frame index
$A=(\hat{i}, \hat{4})$ with $\hat{i}=\hat{1},\hat{2},\hat{3}$.
Note that $\psi=\psi(x), \, q_i= q_i (x)$ and the Killing vector $\partial/\partial \tau$
generates the triholomorphic $U(1)$ symmetry.

One can easily read off the vierbeins from the metric \eq{gh-metric} as
\begin{equation} \label{vE}
E^{\hat{4}} = e^{\psi} (d\tau + q_{i} dx^{i}), \qquad
E^{\hat{i}} = e^{-\psi} dx^{i}
\end{equation}
and
\begin{eqnarray} \label{ivE}
E_{\hat{4}} = e^{-\psi} \frac{\partial}{\partial \tau}, \qquad
E_{\hat{i}} = e^{\psi} \big(\partial_i - q_i \frac{\partial}{\partial \tau} \big).
\end{eqnarray}
Using the torsion-free condition, $T^A=dE^{A} + \omega^{A}_{~B}
\wedge E^{B} =0$, one can calculate the spin connections.
For example, one can get from Eq. (\ref{vE})
\begin{eqnarray*}
dE^{\hat{4}} &=& e^{\psi} \Big( \partial_{i}\psi dx^{i}\wedge d\tau +
\partial_{i}\psi q_{j} dx^{i}\wedge dx^{j}
+ \frac{1}{2} f_{ij} dx^{i}\wedge dx^{j} \Big) \\
&=&-\Big(e^{\psi}\partial_{i}\psi E^{\hat{4}}
+ \frac{1}{2} e^{3\psi} f_{ij} E^{\hat{j}} \Big)\wedge E^{\hat{i}} \\
&=& -\omega_{\hat{4}\hat{i}} \wedge E^{\hat{i}},
\end{eqnarray*}
where $f_{ij} = \partial_{i}q_{j} - \partial_{j}q_{i}$.
Therefore, one can read off
\begin{equation}
\omega_{\hat{4}\hat{i}} = e^{\psi} \partial_{i}\psi E^{\hat{4}} + \frac{1}{2} e^{3\psi} f_{ij}E^{\hat{j}}.
\end{equation}
Similarly, the spin connections and the structure functions can be obtained as follows
\begin{eqnarray} \label{spin-1-form}
\omega_{\hat{4}\hat{i}} &=& e^{\psi} \partial_{i}\psi E^{\hat{4}} + \frac{1}{2} e^{3\psi} f_{ij}E^{\hat{j}}, \nonumber \\
\omega_{\hat{i}\hat{j}} &=& - \frac{1}{2}e^{3\psi} f_{ij} E^{\hat{4}} + e^{\psi} \big(\partial_{i}\psi E^{\hat{j}} -
\partial_{j}\psi E^{\hat{i}} \big), \\
\label{structure-f}
f_{\hat{4}\hat{i}\hat{4}} &=& - \partial_i e^\psi, \qquad
f_{\hat{i}\hat{j}\hat{4}} = e^{3\psi} f_{ij}, \nonumber\\
f_{\hat{4}\hat{i}\hat{j}} &=& 0,  \qquad \qquad f_{\hat{j}\hat{k}\hat{i}} =
\partial_k e^\psi \delta^{\hat{i}}_{\hat{j}} - \partial_j e^\psi \delta^{\hat{i}}_{\hat{k}}.
\end{eqnarray}
Note that we are explicitly discriminating the three-dimensional world and frame indices
as $(i,j,k, \cdots)$ and $(\hat{i},\hat{j},\hat{k}, \cdots)$, respectively.
It is easy to see that the self-duality equation (\ref{sde-spin})
for the spin connection (\ref{spin-1-form}) is reduced to the equation
\begin{equation} \label{abel-self}
\varepsilon_{\hat{i}\hat{j}\hat{k}} \partial_k \psi = \half e^{2\psi} f_{ij} \;\;
\Leftrightarrow \;\;
\nabla V + \nabla \times \vec{q} = 0.
\end{equation}

Using the result (\ref{abel-self}), one can now read off the
self-dual $SU(2)$ gauge fields defined by $\omega_{AB} = A^a
\eta^a_{AB}$:
\begin{eqnarray} \label{su2-gauge}
A^a &=& e^{2\psi} \bar{\eta}^a_{\hat{i}\hat{4}} \partial_i \psi (d\tau + q_{j}
dx^{j}) + \bar{\eta}^a_{\hat{i}\hat{j}} \partial_i \psi dx^{j}
\nonumber \\
&=& e^{\psi} \partial_i \psi \bar{\eta}^a_{\hat{i}A} E^A =
E_{\hat{i}} \psi \bar{\eta}^a_{\hat{i}A} E^A.
\end{eqnarray}
That is, with the notation $E_{\hat{i}} \psi = e^{\psi} \partial_i \psi
\equiv \partial_{\hat{i}} \psi$,
\begin{equation}\label{g-gauge}
A^a_A = \partial_{\hat{i}} \psi \bar{\eta}^a_{\hat{i}A} = \half
 \bar{\eta}^a_{A\hat{i}} \partial_{\hat{i}} \log V.
\end{equation}
It is easy to derive the following relation from Eq. (\ref{abel-self})
\begin{equation} \label{laplace}
e^\psi \partial_i \partial_i e^\psi - 3 \partial_i e^\psi \partial_i
e^\psi = 0.
\end{equation}
Using the above results, one can get the field strengths
for $SU(2)$ gauge fields (\ref{su2-gauge})
\begin{eqnarray} \label{su2-curvature1}
F_{\hat{4}\hat{i}}^a &=& E_{\hat{4}} A_{\hat{i}}^a  - E_{\hat{i}} A_{\hat{4}}^a - 2 \varepsilon^{abc} A^b_{\hat{4}}
A_{\hat{i}}^c + f_{\hat{4}\hat{i}\hat{4}} A_{\hat{4}}^a \nonumber \\
&=&  e^\psi  \partial_i \partial_a e^\psi + 3 \partial_i e^\psi
\partial_a e^\psi - 2 \delta^a_{\hat{i}} \partial_k e^\psi
\partial_k e^\psi, \\
\label{su2-curvature2}
F_{\hat{i}\hat{j}}^a &=& E_{\hat{i}} A_{\hat{j}}^a  - E_{\hat{j}} A_{\hat{i}}^a
- 2 \varepsilon^{abc} A^b_{\hat{i}} A_{\hat{j}}^c
+ f_{\hat{i}\hat{j}\hat{4}} A_{\hat{4}}^a + f_{\hat{i}\hat{j}\hat{k}} A_{\hat{k}}^a \nonumber \\
&=&  e^\psi  \partial_k \Big( \varepsilon_{a\hat{k}\hat{j}} \partial_i e^\psi
 - \varepsilon_{a\hat{k}\hat{i}} \partial_j e^\psi \Big)
 - 4 \varepsilon_{\hat{i}\hat{j}\hat{k}} \partial_k e^\psi \partial_a e^\psi
 + \partial_k e^\psi  \Big( \varepsilon_{a\hat{k}\hat{i}} \partial_j  e^\psi
 - \varepsilon_{a\hat{k}\hat{j}} \partial_i e^\psi \Big).
\end{eqnarray}

Now it is straightforward to check that the above $SU(2)$ field strengths
are self-dual, i.e.
\begin{equation} \label{su2-self-dual}
F_{AB}^a = \frac{1}{2} {\varepsilon_{AB}}^{CD} F_{CD}^a.
\end{equation}
To be specific, one can explicitly see that
\begin{eqnarray} \label{su2-b}
\frac{1}{2} \varepsilon_{\hat{i}\hat{j}\hat{k}} F_{\hat{j}\hat{k}}^a &=&
- e^\psi  \partial_i \partial_a e^\psi - 3 \partial_i e^\psi
\partial_a e^\psi + 2 \delta^a_{\hat{i}} \partial_k e^\psi
\partial_k e^\psi \nonumber \\
&=& F_{\hat{i}\hat{4}}^a,
\end{eqnarray}
where the relation (\ref{laplace}) was used.
In terms of the harmonic function in Eq. (\ref{harmonic-v}),
the above field strength can be represented by
\begin{equation} \label{field-v}
F_{\hat{i}\hat{4}}^a = \frac{1}{2} V^{-2} \partial_i \partial_a V -
\frac{3}{2} V^{-3} \partial_i V \partial_a V
+ \frac{1}{2} \delta^a_{\hat{i}} V^{-3} \partial_k V \partial_k V
\end{equation}
and Eq. (\ref{laplace}) can be written as
\begin{equation} \label{abel-harm-v}
\partial_i \partial_i \log V + \partial_i \log V \partial_i \log V = 0.
\end{equation}
It would be interesting to compare Eq. (\ref{abel-harm-v}) (after the replacement
$\partial_i \to \partial_M$ since the function $V(x)$ does not depend on $\tau$)
with the 't Hooft ansatz $A_\mu^a = \bar{\eta}^a_{\mu\nu} \partial_\nu \log \phi(x)$
for $SU(2)$ multi-instantons (see Eq. (4.60b) in \ct{rajaraman}) satisfying\footnote{Note
that Eq. \eq{laplace} can be represented in terms of frame derivatives as
$ \partial_{\hat{i}} \partial_{\hat{i}} \psi
- 3 \partial_{\hat{i}}\psi \partial_{\hat{i}}\psi = 0$ which also reduces
to the form \eq{thooft-harm-v} with the identification $\psi = - \frac{1}{3} \log \phi$.}
\begin{equation}\label{thooft-harm-v}
\partial_\mu \partial_\mu \log \phi + \partial_\mu \log \phi
\partial_\mu \log \phi = 0.
\end{equation}

Our result here recovers the self-dual gauge fields in \cite{bcc2} (for $H=V$).

\subsection{Taub-NUT metric}

The Taub-NUT metric is the simplest ALF space described by the
Gibbons-Hawking metric \eq{gh-metric} with $\epsilon=1$ and $k=1$.
Using the spherical coordinates, it is given by
\begin{equation} \la{taub-nut}
ds^2 = c_{r}^2 dr^2 + \sum_{i=1}^{3} c_{i}^2 (\sigma^{i})^2
\end{equation}
with the coefficients $c_{1}=c_{2}\ne c_{3}$ given by
\begin{equation}
c_{r}(r) = \frac{1}{2}\sqrt{\frac{r+m}{r-m}},
\quad c_{1}(r)=c_{2}(r)=\frac{1}{2}\sqrt{r^2-m^2}, \quad
c_{3}(r) = m\sqrt{\frac{r-m}{r+m}}.
\end{equation}
The Maurer-Cartan one-forms $\{ \sigma^i \}$ satisfy the following exterior algebra \ct{egh-report}
\begin{equation} \label{mc-s3}
d \sigma^i + \frac{1}{2} \varepsilon^{\hat{i}\hat{j}\hat{k}} \sigma^j \wedge
\sigma^k = 0.
\end{equation}

The vierbein bases are given by
\begin{equation}
E^{\hat{4}} = c_r dr, \qquad E^{\hat{i}} = c_{i}\sigma^{i} \quad (\mathrm{NS}[i]),
\end{equation}
and
\begin{equation}
E_{\hat{4}} =
\frac{1}{c_{r}}\partial_{r}, \qquad E_{\hat{i}}=\frac{1}{c_{i}} \kappa_{i} \quad (\mathrm{NS}[i]),
\end{equation}
where $\kappa_i$ are the basis vectors dual to $\sigma^i$, i.e.
$\langle \sigma^i, \kappa_j \rangle =\delta^i_j$, satisfying
\begin{equation} \label{su2-vector}
[\kappa_i, \kappa_j] = \varepsilon_{\hat{i}\hat{j}\hat{k}} \kappa_k.
\end{equation}
Here we indicate no summation convention for the index $i$ with the notation
$(\mathrm{NS}[i])$. The spin connections read as
\begin{equation} \label{spin-tnut}
\omega_{\hat{i}\hat{4}} = \frac{\partial_{r} c_{i}}{c_{r}} \sigma^{i} \quad (\mathrm{NS}[i]), \qquad
\omega_{\hat{i}\hat{j}} = - \varepsilon_{\hat{i}\hat{j}\hat{k}} \frac{(c_{i}^2+c_{j}^2-c_{k}^2)}{2c_{i}c_{j}}
 \sigma^{k} \quad (\mathrm{NS}[ij]).
\end{equation}
Note that the spin connections in Eq. (\ref{spin-tnut}) are not completely
self-dual, but the anti-self-dual part is simply given by $\omega^{(-)}_{AB}  =
\frac{1}{2} \left( {\omega}_{AB} - \frac{1}{2} {\varepsilon_{AB}}^{CD} \omega_{CD}
\right) = - \bar{\eta}^a_{AB} \frac{\sigma^a}{2}$ and so their curvature tensors identically vanish thanks to Eq. (\ref{mc-s3}).
The curvature tensors are so self-dual, i.e. $R_{AB} = F^a\eta^a_{AB}$, which are given by
\begin{eqnarray} \label{curvature-tnut}
&& R_{\hat{1}\hat{2}} = R_{\hat{3}\hat{4}}= \frac{8m}{(r+m)^3}
\left(E^{\hat{1}} \wedge E^{\hat{2}} + E^{\hat{3}}\wedge E^{\hat{4}}\right), \nonumber \\
&& R_{\hat{1}\hat{4}} = R_{\hat{2}\hat{3}} = - \frac{4m}{(r+m)^3}
\left(E^{\hat{1}} \wedge E^{\hat{4}} + E^{\hat{2}}\wedge E^{\hat{3}}\right), \\
&& R_{\hat{2}\hat{4}} = R_{\hat{3}\hat{1}} = - \frac{4m}{(r+m)^3}
\left(E^{\hat{2}} \wedge E^{\hat{4}} + E^{\hat{3}}\wedge E^{\hat{1}}\right). \nonumber
\end{eqnarray}
The corresponding $SU(2)$ gauge fields can be identified from \eq{spin-tnut} as
\begin{eqnarray} \label{gauge-su2-tnut}
A^{\dot{1}} &\equiv& \frac{1}{2} \Big(\omega_{\hat{1}\hat{4}} + \omega_{\hat{2}\hat{3}}
\Big) = \frac{r-m}{r+m}  \frac{\sigma^1}{2} , \nonumber \\
A^{\dot{2}} &\equiv& \frac{1}{2} \Big( \omega_{\hat{2}\hat{4}} + \omega_{\hat{3}\hat{1}} \Big)
= \frac{r-m}{r+m} \frac{\sigma^2}{2}, \nonumber \\
A^{\dot{3}} &\equiv& \frac{1}{2} \Big( \omega_{\hat{1}\hat{2}} + \omega_{\hat{3}\hat{4}}
\Big) = \Big(-1 + \frac{4m^2}{(r+m)^2} \Big) \frac{\sigma^3}{2}.
\end{eqnarray}
Therefore, the field strength of the $SU(2)$ gauge fields
(\ref{gauge-su2-tnut}) can be calculated to be
\begin{eqnarray} \label{taub-nut-su2}
F &=& dA + A \wedge A \xx
&=& \frac{1}{2}  f^a(r) \eta^a_{AB} E^A \wedge E^B
\end{eqnarray}
with
\begin{equation} \label{su2-3com}
f^{\dot{1}}(r) = - \frac{4m}{(r+m)^3} = f^{\dot{2}}(r), \qquad f^{\dot{3}}(r) =
\frac{8m}{(r+m)^3}.
\end{equation}
Note that the $SU(2)$ field strengths in \eq{taub-nut-su2} are self-dual, i.e. $F = * F$, which, of course, coincide with the curvature tensor (\ref{curvature-tnut}).

Our result here agrees with the self-dual gauge fields in \cite{poyu,kim-yoon1}.

\subsection{Eguchi-Hanson metric}

The Eguchi-Hanson metric \ct{eh-inst} is the simplest ALE space described by the
Gibbons-Hawking metric \eq{gh-metric} with $\epsilon=0$ and $k=2$.
Let us consider the metric given by
\begin{equation} \label{eguchi-hanson}
ds^2 = h^{-2}(r) dr^2 + \frac{r^2}{4}( \sigma_1^2 + \sigma_2^2) +
\frac{r^2}{4} h^2(r) \sigma_3^2
\end{equation}
with the function $h(r) = \sqrt{1 - a^4/r^4}$.
The Maurer-Cartan one-forms $\{ \sigma^i \}$ satisfy the exterior algebra
\begin{equation} \label{mc-eh}
d \sigma^i - \frac{1}{2} \varepsilon^{\hat{i}\hat{j}\hat{k}} \sigma^j \wedge
\sigma^k = 0.
\end{equation}
Note that the sign is different from the Taub-NUT case \eq{mc-s3},
with which the metric \eq{eguchi-hanson} becomes self-dual.
The spin connections are given by Eq. (\ref{spin-tnut}) for $c_{r}
= h^{-1}(r)$, $c_{1}=c_{2}=r/2$, and $c_{3}=r h(r)/2$ and their
components are
\begin{eqnarray}
&& \omega_{\hat{1}\hat{2}} = \omega_{\hat{3}\hat{4}}
= \frac{1}{2} \Big(1+\frac{a^4}{r^4}\Big) \sigma^{3}, \xx
&& \omega_{\hat{1}\hat{4}} = \omega_{\hat{2}\hat{3}}
= \frac{1}{2} \sqrt{1-\frac{a^4}{r^4}}\sigma^{1}, \\
&& \omega_{\hat{2}\hat{4}} = \omega_{\hat{3}\hat{1}} = \frac{1}{2}
\sqrt{1-\frac{a^4}{r^4}} \sigma^{2}, \nonumber
\end{eqnarray}
which are clearly self-dual. The curvature tensors are
straightforwardly computed by
\begin{eqnarray} \la{curvature-eh}
&& R_{\hat{1}\hat{2}} = R_{\hat{3}\hat{4}}= \frac{4a^4}{r^6}
\left(E^{\hat{1}}\wedge E^{\hat{2}}
+ E^{\hat{3}}\wedge E^{\hat{4}} \right), \xx
&& R_{\hat{1}\hat{4}} = R_{\hat{2}\hat{3}}= - \frac{2a^4}{r^6}
\left(E^{\hat{1}}\wedge E^{\hat{4}}
+ E^{\hat{2}}\wedge E^{\hat{3}} \right), \\
&& R_{\hat{2}\hat{4}} = R_{\hat{3}\hat{1}}= - \frac{2a^4}{r^6}
\left(E^{\hat{2}}\wedge E^{\hat{4}}
+ E^{\hat{3}}\wedge E^{\hat{1}} \right). \nonumber
\end{eqnarray}

The self-dual curvature tensors for the Eguchi-Hanson metric
(\ref{eguchi-hanson}) can be determined by $SU(2)$ gauge fields
$A^a = \frac{1}{4} \omega_{AB} \eta^a_{AB} = (f(r) \sigma^1, f(r) \sigma^2, g(r) \sigma^3)$
where
\begin{equation} \label{f-g-eh}
f(r) =\frac{1}{2}\sqrt{1-\frac{a^4}{r^4}}, \qquad g(r)=
\frac{1}{2} \Big( 1+\frac{a^4}{r^4} \Big).
\end{equation}
The corresponding $SU(2)$ field strength coincides with the curvature tensor
$R_{AB} = F^a \eta^a_{AB}$ in \eq{curvature-eh} where
$F^a = dA^a - \varepsilon^{abc} A^b \wedge A^c$ and they are given by
\begin{equation} \label{eh-su2}
F = \frac{1}{2}  f^a(r) \eta^a_{AB} E^A \wedge E^B
\end{equation}
with
\begin{equation} \label{eh-su2-3com}
f^{\dot{1}}(r) = -\frac{2a^4}{r^6}= f^{\dot{2}}(r), \qquad f^{\dot{3}}(r) =
\frac{4a^4}{r^6}.
\end{equation}

Our result here agrees with the self-dual gauge fields in \cite{bcc1,kim-yoon1}.

\subsection{Atiyah-Hitchin metric}

The Atiyah-Hitchin metric \ct{ah-instanton} describes a four-dimensional hyper-K\"ahler manifold with SO(3)
isometry that was introduced to describe the moduli space of $SU(2)$ BPS
monopoles of magnetic charge 2.
Let us consider the Bianchi type IX space \cite{gibbs-pope} which is
locally described by the metric with an $SU(2)$ or $SO(3)$ isometry
group
\begin{equation} \label{bianchi-9}
ds^2 = a_\tau^2 d\tau^2 + \sum_{i=1}^3 a_i^2 (\sigma^i)^2
\end{equation}
where $a_\tau = a_1 a_2 a_3$ and $a_i$'s are functions solely of $\tau$.
The self-dual conditions for all Bianchi IX solutions are given by
the equations
\begin{eqnarray} \label{self-dual-b9}
\frac{1}{a_\tau} \frac{d a_1}{d \tau} &=& \frac{a_2^2 + a_3^2 - a_1^2}{2a_2a_3}
- \alpha_1, \nonumber \\
\frac{1}{a_\tau} \frac{d a_2}{d \tau} &=& \frac{a_3^2 + a_1^2 - a_2^2}{2 a_3a_1}
- \alpha_2, \\
\frac{1}{a_\tau} \frac{d a_3}{d \tau} &=& \frac{a_1^2 + a_2^2 - a_3^2}{2a_1a_2}
- \alpha_3, \nonumber
\end{eqnarray}
where three constant numbers $\alpha_i, \; i = 1, 2, 3$, satisfy
$\alpha_i \alpha_j = \varepsilon_{\hat{i}\hat{j}\hat{k}} \alpha_k$. Choosing
$(\alpha_1, \alpha_2, \alpha_3) = (1, 1, 1)$ will lead to the
Atiyah-Hitchin metric \cite{ah-instanton} while $(\alpha_1,
\alpha_2, \alpha_3) = (0, 0, 0)$ yields the Eguchi-Hanson type I or
II metric \cite{eh-inst}.

Identify the vierbein basis from the metric (\ref{bianchi-9})
\begin{equation} \label{ah-1-form}
\{E^{\hat{i}}, E^{\hat{4}} \}
= \{ a_i \sigma^i,  a_\tau d \tau \}, \qquad
\{E_{\hat{i}}, E_{\hat{4}} \}
= \{ a^{-1}_i \kappa_i, a^{-1}_\tau \frac{\partial}{\partial \tau} \}
\end{equation}
without summation convention for the index $i$.
The left-invariant 1-forms $\{ \sigma^i \}$ on $\mathbf{S}^3$ satisfy
the exterior algebra \eq{mc-eh} and the dual basis vectors $\{ \kappa_i \}$ satisfy
the Lie algebra $[\kappa_i, \kappa_j] = - \varepsilon_{\hat{i}\hat{j}\hat{k}} \kappa_k$.
Note that the metric \eq{bianchi-9} has the same structure as the Taub-NUT
metric \eq{taub-nut}. Therefore, the spin connections
also have the same structure as follows
\begin{equation}    \label{ah-spin}
\omega_{\hat{i}\hat{4}} = \frac{a'_i}{a_\tau} \sigma^i  \quad (\mathrm{NS}[i]),
\qquad \omega_{\hat{i}\hat{j}} = \varepsilon_{\hat{i}\hat{j}\hat{k}}
 \frac{a_i^2 + a_j^2 - a_k^2}{2 a_i a_j} \sigma^k \quad (\mathrm{NS}[ij]),
\end{equation}
where the prime means the derivative with respect to $\tau$.
Note that the spin connections in (\ref{ah-spin}) are not self-dual
in general. One can check using Eq. \eq{self-dual-b9} that
the spin connections in Eq. (\ref{ah-spin}) satisfy the following relation
\begin{equation}\label{ah-spin-sde}
    \frac{1}{4} \bar{\eta}^a_{AB} \omega_{AB} = \frac{1}{2} \Big(
    - \omega_{a\hat{4}} + \frac{1}{2}
    \varepsilon_{a\hat{j}\hat{k}} \omega_{\hat{j}\hat{k}} \Big)
    = \frac{1}{2} \alpha_{a} \sigma^{a} \quad (\mathrm{NS}[a]).
\end{equation}
Therefore, they become self-dual only when $(\alpha_1, \alpha_2, \alpha_3) = (0, 0, 0)$
which was completely solved. (See Eq. (4.23)
in \cite{gibbs-pope} for the exact solution.) But the curvature tensors
will be self-dual, i.e. $F_{MN}^{(-)} = 0$ in Eq. \eq{dec-curv},
because the curvature tensor of the anti-self-dual spin connections
in \eq{ah-spin-sde} identically vanishes due to Eq. \eq{mc-eh}.

Let us define $SU(2)$ gauge fields as follows
\begin{equation}\label{ah-su2}
    A^a \equiv \frac{1}{4} \eta^a_{AB} \omega_{AB} = \omega_{a\hat{4}}
    + \frac{1}{2} \alpha_a \sigma^a \quad (\mathrm{NS}[a]).
\end{equation}
Our previous result \eq{ym-instanton} implies that the field strengths
$F^a = dA^a - \varepsilon^{abc} A^b \wedge A^c$ defined by the $SU(2)$ gauge fields
in (\ref{ah-su2}) are necessarily self-dual. Now we will show that it is the case.
It is straightforward to calculate the $SU(2)$ field strength
\begin{eqnarray}\label{ah-su2-curv}
F^a &=& \Big(\frac{a'_a}{a_{\tau}} \Big)' d\tau \wedge \sigma^{a}
+ \varepsilon^{abc} \Big(\frac{a'_a}{2a_{\tau}} - \frac{a'_{b}a'_{c}}{a^{2}_{\tau}} - \frac{a'_{b}\alpha_{c}}{a_{\tau}} \Big)
\sigma^{b} \wedge \sigma^{c} \quad (\mathrm{NS}[a]) \xx
&=& \widetilde{a}'_{a} d\tau \wedge \sigma^{a} + \varepsilon^{abc}
\Big(\frac{\widetilde{a}_a}{2} - \widetilde{a}_{b}\widetilde{a}_{c} -
\widetilde{a}_{b}\alpha_{c} \Big) \sigma^{b} \wedge \sigma^{c} \quad (\mathrm{NS}[a]),
\end{eqnarray}
where $\widetilde{a}_a \equiv a'_a/a_{\tau}$. Using the identity \ct{hana-piol}
derived from Eq. \eq{self-dual-b9},
\begin{equation}\label{ah-identity}
    \frac{\widetilde{a}'_1}{a_{1}a_{\tau}} = - \frac{\widetilde{a}_{1}
    - 2 \widetilde{a}_{2} \widetilde{a}_{3} - \widetilde{a}_{2}\alpha_{3}
    - \widetilde{a}_{3}\alpha_{2}}{a_{2}a_{3}}, \quad \mathrm{etc},
\end{equation}
we see that the field strength \eq{ah-su2-curv} has the correct self-dual structure, i.e.
\begin{eqnarray}\label{ah-sdcurv}
F &=& dA + A \wedge A \xx
&=& \frac{1}{2} f^{a}(\tau) \eta^a_{AB} E^A \wedge E^B \xx
&=& - \frac{\widetilde{a}'_a}{2 a_{a}a_{\tau}} \eta^a_{AB} E^A \wedge E^B.
\end{eqnarray}

The self-dual gauge fields in Eqs. \eq{ah-su2} and \eq{ah-su2-curv}
describe a Yang-Mills instanton on the Atiyah-Hitchin space
and it consists of a new solution to the extent of our knowledge.

\subsection{Real heaven}

The real heaven metric \cite{real-heaven} describes four dimensional
hyper-K\"ahler manifolds with a rotational Killing symmetry which is
also completely determined by one real scalar field. The metric is
given by
\begin{eqnarray} \label{heaven}
ds^2 &=& (\partial_3 \psi)^{-1}(d\tau +  q_\alpha d x^\alpha)^2 + (\partial_3
\psi) ( e^\psi dx^\alpha dx^\alpha + dx^3 dx^3) \xx
&\equiv& e^{-2\phi_{4}} (d\tau + q_\alpha dx^{\alpha})^2 + e^{2\phi_i} dx^{i}dx^{i}
\end{eqnarray}
where $q_\alpha = - \varepsilon^{\alpha\beta} \partial_\beta \psi, \; (\alpha=1,2)$ and the
function $\psi(x)$ is independent of $\tau$ and satisfy the
three-dimensional continual Toda equation
\begin{equation} \label{toda}
(\partial_1^2 + \partial_2^2) \psi + \partial_3^2 e^\psi = 0.
\end{equation}
The rotational Killing vector is given by $c_\alpha \partial_\alpha \psi
\partial/\partial \tau$ with constants $c_\alpha$.

We identify the vierbein vectors as
\begin{equation}
E^{{\hat{i}}}=e^{\phi_i} dx^i \quad (\mathrm{NS}[i]), \qquad E^{{\hat{4}}} = e^{-\phi_4}(d\tau+q_\alpha dx^{\alpha}).
\end{equation}
where
\begin{equation}
e^{2\phi_1} =  e^{2\phi_2} = \partial_{3} \psi e^{\psi}, \qquad e^{2\phi_3} =
e^{2\phi_4} = \partial_{3} \psi.
\end{equation}
From the torsion-free equation $dE^{A}+\omega^{A}_{~B}\wedge
E^{B}=0$, we get
\begin{eqnarray} \la{spin-heaven}
&& \omega_{\hat{i}\hat{4}} =
- e^{\phi_4 - \phi_i} \partial_{i} e^{-\phi_4} E^{{\hat{4}}}
- \frac{1}{2} e^{-\phi_4 - \phi_i - \phi_j} f_{ij}
E^{{\hat{j}}} \quad (\mathrm{NS}[i]), \nonumber \\
&& \omega_{\hat{i}\hat{j}} = -\frac{1}{2} e^{-\phi_4 - \phi_i - \phi_j} f_{ij} E^{{\hat{4}}}
+ e^{-\phi_i - \phi_j} \big( \partial_{j} e^{\phi_i} E^{{\hat{i}}}  -
\partial_{i} e^{\phi_j} E^{{\hat{j}}} \big) \quad (\mathrm{NS}[ij]),
\end{eqnarray}
where $f_{ij} = \partial_i q_j - \partial_j q_i$ with
$q_i \equiv - \varepsilon^{3ij} \partial_j \psi$.

It is straightforward to check that the self dual relations,
$\omega_{{\hat{3}}{\hat{1}}} = \omega_{{\hat{2}}{\hat{4}}}$ and $\omega_{{\hat{2}}{\hat{3}}}=\omega_{{\hat{1}}{\hat{4}}}$, are satisfied
if and only if the continual Toda equation \eq{toda} is satisfied.
However, the relation $\omega_{{\hat{1}}{\hat{2}}} =  \omega_{{\hat{3}}{\hat{4}}}$ is not satisfied.
In order to cure this mismatch, first note that we can perform the
local frame rotation (\ref{frame-rotation}) as follows
\begin{eqnarray} \label{rh-so4-frame}
\widetilde{E}^A &=& {\Lambda^A}_B E^B \nonumber \\
&=& \left(
                    \begin{array}{cccc}
                      1 & 0 & 0 & 0 \\
                      0 & 1 & 0 & 0 \\
                      0 & 0 & \cos \frac{\tau}{2} & -\sin \frac{\tau}{2} \\
                      0 & 0 & \sin \frac{\tau}{2} & \cos \frac{\tau}{2} \\
                    \end{array}
                  \right)
\left(
  \begin{array}{c}
    E^{\hat{1}} \\
    E^{\hat{2}} \\
    E^{\hat{3}} \\
    E^{\hat{4}} \\
  \end{array}
\right).
\end{eqnarray}
The spin connections also transform according to Eq.(\ref{spin-so4})
\begin{equation} \label{rh-so4-spin}
{\widetilde{\omega}^A}_{~~B} = {\Lambda^A}_C {{\omega}^C}_D {\Lambda^{-1
D}}_B + {\Lambda^A}_C {(d \Lambda^{-1})^C}_B
\end{equation}
where
\begin{equation} \label{rh-spin-inhom}
{\Lambda^A}_C {(d \Lambda^{-1})^C}_B = \frac{1}{2} \left(
                    \begin{array}{cccc}
                      0 & 0 & 0 & 0 \\
                      0 & 0 & 0 & 0 \\
                      0 & 0 & 0 & 1 \\
                      0 & 0 & -1 & 0 \\
                    \end{array}
                  \right) d \tau
\end{equation}
and $d\tau= \big( - q_\alpha e^{-\phi_\alpha} E^{\hat{\alpha}}
+ e^{\phi_4} E^{\hat{4}} \big)$.
Note that the frame rotation (\ref{rh-so4-frame}) affects the self-duality condition
only for ${\widetilde{\omega}}_{{\hat{3}}{\hat{4}}} = {\omega}_{\hat{3}\hat{4}}
+ \frac{1}{2} d\tau$ due to the inhomogeneous term (\ref{rh-spin-inhom}).
In other words, ${\widetilde{\omega}}_{{\hat{3}}{\hat{1}}} =  {\widetilde{\omega}}_{{\hat{2}}{\hat{4}}}$ and
${\widetilde{\omega}}_{{\hat{2}}{\hat{3}}} = {\widetilde{\omega}}_{{\hat{1}}{\hat{4}}}$ are
automatically satisfied thanks to the previous relations. Now it is
straightforward to check that ${\widetilde{\omega}}_{{\hat{1}}{\hat{2}}} = {\omega}_{{\hat{1}}{\hat{2}}} =
{\widetilde{\omega}}_{{\hat{3}}{\hat{4}}} =  \big(\omega_{{\hat{3}}{\hat{4}}}
- \frac{1}{2} q_\alpha e^{-\phi_\alpha} E^{\hat{\alpha}}
+ \frac{1}{2} e^{\phi_4} E^{\hat{4}} \big)$. Therefore, the spin connections
in \eq{rh-so4-spin} become self-dual.

If one introduces $SU(2)$ gauge fields by
\begin{equation}\label{heaven-su2}
    A^a \equiv \frac{1}{4} \eta^a_{AB} \widetilde{\omega}_{AB} =
    \widetilde{\omega}_{a\hat{4}}
    = \omega_{a\hat{4}} + \frac{1}{2} \delta^a_{\hat{3}} d\tau,
\end{equation}
the corresponding field strengths, $F^a = dA^a - \varepsilon^{abc} A^b \wedge A^c$,
should be self-dual according to the general
result \eq{ym-instanton}. This can also be proved by using the relation \eq{eta-o4-algebra}
which leads to the following result
\begin{equation}\label{su2-sd-proof}
    F^a = \frac{1}{4} \eta^a_{AB} \Big( d\widetilde{\omega}_{AB} +
    \widetilde{\omega}_{AC} \wedge \widetilde{\omega}_{CB} \Big) =
    \frac{1}{4} \eta^a_{AB} \widetilde{R}_{AB}.
\end{equation}
Hence the self-duality of $F^a$ results from
the self-dual curvature tensors $\widetilde{R}_{AB}$.
Or one can check it by a straightforward calculation
using Eqs. \eq{spin-heaven} and \eq{toda} though rather tedious.

The self-dual gauge fields in Eq. \eq{heaven-su2}
describe a Yang-Mills instanton on the real heaven \eq{heaven}, which is
a new solution to the extent of our knowledge.

\subsection{Euclidean Schwarzschild solution}

The Euclidean Schwarzschild metric \ct{gi-hawking} was constructed by
the Wick rotation of the Schwarzschild black-hole solution.
It is not a gravitational instanton (not a hyper-K\"ahler manifold) although
it is a Ricci-flat manifold. The metric takes the form
\begin{equation}\label{e-sbh}
    ds^2 = \Big(1 - \frac{2m}{r} \Big) d\tau^{2} + \Big(1 - \frac{2m}{r} \Big)^{-1} dr^{2}
+ r^{2} (d \theta^{2} + \sin \theta^2 d\phi^2).
\end{equation}
The radial coordinate is constrained by $r \geq 2m$ and the time coordinate $\tau$
is an angular variable with period $8\pi m$. Hence this solution
has the topology $\mathbb{R}^{2} \times \mathbb{S}^{2}$.

After defining the vierbein basis $(E^{\hat{1}}= h(r)^{-1}dr,
E^{\hat{2}}= r d\theta, E^{\hat{3}}= r\sin \theta d\phi, E^{\hat{4}}= h(r)d\tau)$,
it is easy to compute spin connections:
\begin{eqnarray} \la{spin-esbh}
  && \omega_{\hat{1}\hat{2}} = - h d\theta, \quad
  \omega_{\hat{1}\hat{3}} = - h \sin \theta d\phi,
  \quad \omega_{\hat{2}\hat{3}} = - \cos \theta d\phi,  \xx
  && \omega_{\hat{1}\hat{4}} = - \frac{1}{2} (h^2)' d\tau, \quad
  \omega_{\hat{2}\hat{4}}= \omega_{\hat{3}\hat{4}} = 0,
\end{eqnarray}
where $h(r) = \sqrt{1-\frac{2m}{r}}$.
The corresponding curvature tensors are given by
\begin{eqnarray} \la{curv-esbh}
&& R_{\hat{1}\hat{2}} = - \frac{m}{r^{3}} E^{\hat{1}} \wedge E^{\hat{2}},
\quad R_{\hat{1}\hat{3}} = - \frac{m}{r^{3}} E^{\hat{1}} \wedge E^{\hat{3}},  \quad R_{\hat{1}\hat{4}} = \frac{2m}{r^{3}} E^{\hat{1}} \wedge E^{\hat{4}}, \xx
&&  R_{\hat{2}\hat{3}} = \frac{2m}{r^{3}} E^{\hat{2}} \wedge E^{\hat{3}},
\quad R_{\hat{2}\hat{4}} = - \frac{m}{r^{3}} E^{\hat{2}} \wedge E^{\hat{4}},
\quad R_{\hat{3}\hat{4}} = - \frac{m}{r^{3}} E^{\hat{3}} \wedge E^{\hat{4}},
\end{eqnarray}
which are not self-dual anymore although they are Ricci-flat, i.e.,
$R_{AB} \equiv R_{ACBC} = 0$.

Because the spin connections in Eq. \eq{spin-esbh} are neither self-dual nor anti-self-dual,
we can consider both type of $SU(2)$ gauge fields defined by
\begin{equation}\label{esbh-su2}
    A^{(\pm)a} \equiv \frac{1}{4} \eta^{(\pm)a}_{AB} \omega_{AB}.
\end{equation}
The field strengths, $F^{(\pm)a} = dA^{(\pm)a} - \varepsilon^{abc} A^{(\pm)b} \wedge A^{(\pm)c}$,
should be either self-dual (for the $+$ sign) or anti-self-dual (for the $-$ sign)
because we get the following result
\begin{equation}\label{su2-esbh-proof}
    F^{(\pm)a} = \frac{1}{4} \eta^{(\pm)a}_{AB} \Big( d\omega_{AB} +
    \omega_{AC} \wedge \omega_{CB} \Big) =
    \frac{1}{4} \eta^{(\pm)a}_{AB} R_{AB},
\end{equation}
which can be derived by using the relation \eq{eta-o4-algebra}.
According to the general result \eq{ym-instanton}, the $SU(2)$ gauge fields in
Eq. \eq{esbh-su2} automatically satisfy the self-duality equation \eq{su2-instanton}
where the background geometry is given by the metric \eq{e-sbh}.
Therefore, the solution \eq{esbh-su2} indeed describes an $SU(2)$ Yang-Mills (anti-)instanton
on the space \eq{e-sbh}.

The solution \eq{esbh-su2} was originally found by Charap and Duff \ct{duff}.
The reason for the revival here is that the solution \eq{esbh-su2} exposes an interesting structure for a Ricci-flat manifold. According to the decomposition \eq{dec-spin} and \eq{dec-curv}, we see that the Euclidean Schwarzschild metric \eq{e-sbh} describes
the sum of an $SU(2)_L$ instanton and an $SU(2)_R$ anti-instanton.
Therefore, an interesting question is whether this kind of feature is generic or not.
Remarkably it can be shown \ct{our-next} that any Einstein manifold satisfying
$R_{AB} = \Lambda \delta_{AB}$ for either $\Lambda=0$ or $\Lambda \neq 0$
always arises as the sum of $SU(2)_L$ instantons and $SU(2)_R$ anti-instantons.

\section{Topological Invariants}

The correspondence between gravitational and Yang-Mills instantons now
raises an intriguing question about topological invariants in
gravity and gauge theories. In the gravity side, there are two topological invariants
associated with the Atiyah-Patodi-Singer index theorem for an elliptic
complex in four dimensions \cite{egh-report}, namely the Euler
characteristic $\chi(M)$ and the Hirzebruch signature $\tau(M)$,
which can be expressed as integrals of the curvature of a four dimensional metric
while, in the gauge theory side, there is a unique topological invariant
up to a boundary term given by the Chern class of gauge bundle. Thus a natural
question is how the two kinds of topological invariants for self-dual
four manifolds can be related to the Chern class of instanton bundle.
In particular, the two topological invariants for gravitational instantons
should be related to each other, in other words,
\be \la{top-relation}
a \chi(M) + b \tau(M) = c, \qquad a,b,c \in \mathbb{Z},
\ee
because there is only a unique topological invariant $c_2(E)$, the second Chern class, for Yang-Mills instantons.

The topologically inequivalent sector of instanton solutions is defined
by the homotopy class of a map from a three sphere at asymptotic infinity
into the gauge group $G=SU(2)$
\begin{equation}\label{homotopy-map}
    f: \mathbb{S}^3 \to SU(2)
\end{equation}
and the topological charge is defined by an element of the homotopy
group $\pi_3 (SU(2)) = \mathbb{Z}$. Viewed the spin connections
in Eq. \eq{spin-so4} as gauge fields in $G= O(4) = SU(2)_L \times SU(2)_R$,
one may also classify the topological sectors of the $O(4)$ gauge fields
in Eq. \eq{spin-sd-asd} by the homotopy class of the map
\begin{equation}\label{homotopy-spin-map}
    f: \mathbb{S}^3 \to O(4) = SU(2)_L \times SU(2)_R.
\end{equation}
Hence the homotopy group of $O(4)$ in the gravity theory is isomorphic to
two copies of the additive group of integers
\begin{equation}\label{gravity-homotopy}
 \pi_3 (O(4)) \approx \pi_3 (SU(2)_L \times SU(2)_R)
 \approx \mathbb{Z} \oplus \mathbb{Z}.
\end{equation}
Consequently, there are two independent gravitational topological
charges \ct{egh-report}, i.e., the Euler characteristic $\chi(M)$ and the Hirzebruch signature $\tau(M)$.

The Euler number $\chi(M)$ for the de Rham complex and the signature $\tau(M)$
for the Hirzebruch signature complex are, respectively, defined by\footnote{\la{orientation}Note
that our definition is different in signs of boundary terms from that in \ct{egh-report}
because we choose the orientation $d^3 x \wedge d\tau  = -d\tau \wedge d^3 x$
to be positive and the $\tau$-direction to be normal to the boundary $\partial M$
while the orientation $d\tau \wedge d^3 x$
was chosen to be positive in \ct{egh-report}.}
\begin{eqnarray} \label{euler}
\chi(M) & = & \frac{1}{32 \pi^2} \int_{M }\varepsilon^{ABCD}
R_{AB} \wedge R_{CD} \xx
&& + \frac{1}{16 \pi^2} \int_{\partial M}
\varepsilon^{ABCD} \Big( \theta_{AB} \wedge R_{CD}
- \frac{2}{3} \theta_{AB} \wedge \theta_{CE} \wedge \theta_{ED} \Big), \\
\label{signature}
\tau(M) &=& - \frac{1}{24 \pi^2} \int_{M} \mathrm{Tr} R \wedge R
 - \frac{1}{24 \pi^2} \int_{\partial M}
\mathrm{Tr} \theta \wedge R + \eta_S(\partial M),
\end{eqnarray}
where $\theta_{AB}$ is the second fundamental form of the boundary
$\partial M$. It is defined by
\begin{equation} \label{2-f-form}
\theta_{AB} = \omega_{AB} - \omega_{0AB},
\end{equation}
where $\omega_{AB}$ are the actual connection 1-forms and
$\omega_{0AB}$ are the connection 1-forms if the
metric were locally a product form near the boundary \ct{egh-report}.
The connection 1-form $\omega_{0AB}$ will have only tangential
components on $\partial M$ and so the second fundamental form
$\theta_{AB}$ will have only normal components on $\partial M$.
And $\eta_S(\partial M)$ is the $\eta$-function given by the eigenvalues
of a signature operator defined over $\partial M$ and depends only
on the metric on $\partial M$ \cite{egh-report}.
The topological invariants are also related to
nuts (isolated points) and bolts (two surfaces),
which are the fixed points of the action of
one parameter isometry groups of gravitational instantons \ct{gh-cmp}.

We have verified in the previous sections that, for gravitational instantons,
one of the $SU(2)$ factors in \eq{homotopy-spin-map} completely decouples
from the theory. Therefore, the topological classification of (anti-)self-dual
spin connections will essentially be the same as Eq. \eq{homotopy-map} in the gauge theory.
That is the reason why we expect the relation \eq{top-relation}
for the topological invariants in Eqs. \eq{euler} and \eq{signature}.
Now we will confirm the relation \eq{top-relation} explicitly determining the coefficients.

Since $\theta_{AB}$ in Eq. (\ref{2-f-form}) are antisymmetric on the
$AB$ index pair, we will decompose them into a self-dual part and an
anti-self-dual part according to Eq. (\ref{spin-sd-asd})
\begin{equation} \label{2-f-sd-asd}
\theta_{AB} \equiv a^{(+)a} \eta^a_{AB} + a^{(-)a}
\bar{\eta}^a_{AB}.
\end{equation}
We take the normal to the boundary to be $(A=\hat{4})$-direction and
so we have $\theta_{\hat{i}\hat{j}}=0$. It
is then straightforward to express the topological invariants
in terms of $SU(2)$ gauge fields using the
decompositions (\ref{spin-sd-asd}), (\ref{curvature-sd-asd}) and
(\ref{2-f-sd-asd}):
\begin{eqnarray} \label{euler-gauge}
\chi(M) &=& \frac{1}{4 \pi^2} \int_M  \Big( F^{(+)a}  \wedge
F^{(+)a}  - F^{(-)a}  \wedge F^{(-)a} \Big) \nonumber \\
&& + \frac{1}{4 \pi^2} \int_{\partial M}
\Big( a^{(+)a} - a^{(-)a} \Big) \wedge
\Big( F^{(+)a} + F^{(-)a} \Big) \nonumber \\
&& + \frac{1}{12 \pi^2} \int_{\partial M}
\varepsilon^{abc} \Big( a^{(+)a} - a^{(-)a} \Big) \wedge
\Big( a^{(+)b} - a^{(-)b} \Big)
\wedge \Big( a^{(+)c} - a^{(-)c} \Big) \nonumber \\
&=&  \frac{1}{16 \pi^2} \int_M  \sqrt{g} \varepsilon^{ABCD} \Big( F_{AB}^{(+)a}
F_{CD}^{(+)a} - F_{AB}^{(-)a} F_{CD}^{(-)a} \Big)  d^4x  \nonumber \\
&& + \frac{1}{8 \pi^2} \int_{\partial M} \sqrt{h} \varepsilon^{\hat{i}\hat{j}\hat{k}}
\Big( a_{\hat{i}}^{(+)a}- a_{\hat{i}}^{(-)a} \Big) \Big(F_{\hat{j}\hat{k}}^{(+)a}
+ F_{\hat{j}\hat{k}}^{(-)a} \Big) d^3x \nonumber \\
&& + \frac{1}{12 \pi^2} \int_{\partial M} \sqrt{h} \varepsilon^{abc} \varepsilon^{\hat{i}\hat{j}\hat{k}}
\Big( a^{(+)a}_{\hat{i}} - a^{(-)a}_{\hat{i}} \Big)
\Big( a^{(+)b}_{\hat{j}} - a^{(-)b}_{\hat{j}} \Big)
\Big( a^{(+)c}_{\hat{k}} - a^{(-)c}_{\hat{k}} \Big) d^3x, \\
\label{signature-gauge}
\tau(M) &=& \frac{1}{6 \pi^2} \int_{M} \Big( F^{(+)a} \wedge F^{(+)a} +
F^{(-)a}  \wedge F^{(-)a} \Big) \nonumber \\
&& + \frac{1}{12 \pi^2} \int_{\partial M} \Big( a^{(+)a} - a^{(-)a} \Big) \wedge
\Big( F^{(+)a} - F^{(-)a} \Big) + \eta_S(\partial M) \nonumber \\
&=& \frac{1}{24 \pi^2} \int_M  \sqrt{g} \varepsilon^{ABCD} \Big(
F_{AB}^{(+)a} F_{CD}^{(+)a} + F_{AB}^{(-)a} F_{CD}^{(-)a} \Big)
d^4x \nonumber \\
&& + \frac{1}{24 \pi^2} \int_{\partial M} \sqrt{h} \varepsilon^{\hat{i}\hat{j}\hat{k}}
\Big( a^{(+)a}_{\hat{i}} - a^{(-)a}_{\hat{i}} \Big)
\Big( F^{(+)a}_{\hat{j}\hat{k}} - F^{(-)b}_{\hat{j}\hat{k}} \Big) d^3x
+ \eta_S(\partial M),
\end{eqnarray}
where we defined the volume forms as $E^{\hat{1}} \wedge E^{\hat{2}} \wedge
E^{\hat{3}} \wedge E^{\hat{4}} \equiv \sqrt{g} d^4 x$ and
$E^{\hat{1}} \wedge E^{\hat{2}} \wedge E^{\hat{3}}|_{\partial M}
\equiv \sqrt{h} d^3 x$.

An interesting pattern appears in the topological invariants.
First consider a compact Einstein manifold without boundary, i.e.
$\partial M = 0$. It turns out \ct{our-next} that $F^{(+)a}$ and
$F^{(-)a}$ are self-dual and anti-self-dual instantons, respectively.
Then we see that the Euler number $\chi(M) = \chi^+(M) + \chi^-(M)$
does not distinguish self-dual and anti-self-dual instantons
since both contribute with equal sign while the Hirzebruch signature
$\tau(M) = \tau^+(M) - \tau^-(M)$ distinguishes self-dual and
anti-self-dual instantons. Based on the observation,
we can draw general properties about 4-dimensional compact Einstein
manifolds where all boundary terms vanish. As we mentioned above,
the Euler number $\chi(M)$ gets equal sign contributions from
self-dual and anti-self-dual gauge fields while the Hirzebruch
signature $\tau(M)$ is not the case. Thus we see that $\chi(M) \geq
0$ with the equality only if $M$ is flat. This is the Berger's
result \ct{egh-report}. We can further refine the Berger's result by looking at the expressions
(\ref{euler-gauge}) and (\ref{signature-gauge}):
\begin{equation}\label{es-top}
  \chi(M) - \frac{3}{2} \tau(M) = - \frac{1}{2 \pi^2} \int_M F^{(-)a}
  \wedge F^{(-)a} \geq 0
\end{equation}
because $F^{(-)}$ describes $SU(2)$ anti-instantons. The inequality \eq{es-top}
will be saturated if and only if a compact four-manifold is half-flat, i.e.
$F^{(-)a} = 0$. In the result, we get a general relation
\be \la{hitchin-bound}
\chi(M) \geq \frac{3}{2} |\tau(M)|
\ee
where the bound is saturated only for $\mathbb{T}^4$ and $K3$ surface, which are
compact self-dual four-manifolds as either trivial or nontrivial
gravitational instantons. This result is known as the Hitchin-Thorpe
inequality \cite{egh-report}.

For noncompact manifolds, there are additional boundary terms as shown in
(\ref{euler-gauge}) and (\ref{signature-gauge}) which are not separated into the self-dual and anti-self-dual parts unlike as the volume terms. In particular,
the eta-invariant $\eta_S(\partial M)$ for $k$ self-dual
gravitational instantons \cite{gibb-perry} is given by
\begin{equation} \label{index-eta}
\eta_S(\partial M) = - \frac{2 \epsilon}{3 k} + \frac{(k-1)(k-2)}{3k}
\end{equation}
where $\epsilon = 0$ for ALE boundary conditions and $\epsilon = 1$
for ALF boundary conditions. Because the topological invariants for a noncompact
manifold with boundary have nontrivial boundary corrections, it is not easy to
demonstrate the relation \eq{top-relation} although such a relation should
exist for general half-flat manifolds. But, one may infer by investigating
known examples so far that the following relation
\begin{equation}\label{top-rel-nc}
\chi(M) = |\tau (M)| + 1
\end{equation}
would be satisfied for noncompact gravitational instantons.
It turns out \ct{gh-cmp,gpr-index,gibb-perry} that ALE instantons including all ADE series and ALF instantons of AD series satisfy the relation \eq{top-rel-nc}.\footnote{$A_{k-1}$ ALE ($\epsilon =0$) and ALF ($\epsilon=1$) instantons are described by the Gibbons-Hawking metric \eq{gh-metric} and $D_0$ ALF instantons are described by the Atiyah-Hitchin metric \eq{bianchi-9}. Especially, Kronheimer obtained the explicit construction of the ALE manifolds as hyper-K\"ahler quotients \ct{kronheimer} which heavily relies on the algebraic structure of the Kleinian groups $\Gamma$ and the crucial identification between
the Hirzebruch signature $\tau(M)$ and the number of conjugacy classes
of the finite group $\Gamma$. See the Table 2 in \ct{italy2} for the relation
\eq{top-rel-nc} of all ALE manifolds. See also the Table D.1 in \ct{egh-report}.}

Therefore, the evidence for the relation \eq{top-rel-nc} is overwhelming.
Since we believe that the relation \eq{top-relation} will be generic independently
of asymptotic boundary conditions and topology,
we conjecture that the relation \eq{top-rel-nc}
will be true for general noncompact gravitational instantons.
It may be proved by showing the following identity for gravitational instantons, e.g.,
with $F^{(-)a}=0$ and so taking the self-dual gauge $A^{(-)a}=0$:
\begin{eqnarray} \label{conjecture}
\chi(M) - \tau(M) &=& \frac{1}{12 \pi^2} \int_M  F^{(+)a}  \wedge
F^{(+)a} + \frac{1}{6 \pi^2} \int_{\partial M} a^{(+)a} \wedge F^{(+)a} \nonumber \\
&& + \frac{1}{12 \pi^2} \int_{\partial M}
\varepsilon^{abc} a^{(+)a} \wedge a^{(+)b} \wedge a^{(+)c} - \eta_S(\partial M) \xx
&=& 1.
\end{eqnarray}
Indeed, for ALE and ALF spaces, one can derive the relation $\chi(M) - \tau(M) = 1 - 4 I_{\frac{1}{2}}(S_\pm, D)$ using Eqs. (12), (13), (14) and (20) in \ct{gpr-index}.
If $M$ has a spin structure, the index of the Dirac operator, $I_{\frac{1}{2}}(S_\pm, D)$,
must identically vanish \ct{gibbs-pope}, and thus we confirm the above identity.
For general cases, we do not know how to rigorously prove the above identity
and so we leave it as our conjecture.

The topological invariant in $SU(2)$ gauge theory is given by the
second Chern number
\begin{equation} \label{top-num-su2}
k = \frac{1}{16 \pi^2} \int_M F_{YM}^a \wedge F_{YM}^a
\end{equation}
where $F_{YM}^a = d A_{YM}^a + \frac{1}{2} \varepsilon^{abc}
A_{YM}^b \wedge A_{YM}^c$. Note that the $SU(2)$ field strength
coming from the spin connections is given by $F_{G}^a = d A_{G}^a -
\varepsilon^{abc} A_{G}^b \wedge A_{G}^c$. So they are related by
$A_{YM}^a = - 2 A_{G}^a$ and $F_{YM}^a = - 2 F_{G}^a$ \ct{kim-yoon1}. Taking this
factor into account, one can see that the Chern number (\ref{top-num-su2}) has
the same normalization factor as the Euler number in Eq. (\ref{euler-gauge}),
i.e.,
\begin{equation} \label{top-num-gi}
k = \frac{1}{4 \pi^2} \int_M F_{G}^a \wedge F_{G}^a.
\end{equation}
This fact provides us an interesting insight why the instanton number
(\ref{top-num-gi}) for $SU(2)$ instantons satisfying \eq{su2-instanton} is
not necessarily integer-valued \cite{bcc1,bcc2}. Note that the Euler number \eq{euler}
as well as the signature \eq{signature} are all integer-valued.
Therefore, if there is a nontrivial boundary correction in
the Euler number \eq{euler-gauge},  the instanton number \eq{top-num-gi}
will not be an integer, i.e., a fractional number in general.
We will illustrate it with explicit examples.

\subsection{Taub-NUT space}

For the product metric
\begin{equation}
ds^2 = \frac{1}{4} \frac{r_0+m}{r_0-m} dr^2 + \frac{1}{4}(r_0^2-m^2)
(\sigma_{1}^2 + \sigma_2^2) +  m^2 \frac{r_0-m}{r_0+m} \sigma_3^2,
\end{equation}
the spin connections are given by
\begin{equation}
(\omega_0)_{\hat{i}\hat{4}} = 0, \qquad  (\omega_0)_{\hat{i}\hat{j}}=
\omega_{\hat{i}\hat{j}}( r = r_0).
\end{equation}
Hence the second fundamental form at the boundary $r = r_0$ is
\begin{equation}
\theta_{\hat{i}\hat{4}} = \omega_{\hat{i}\hat{4}}( r = r_0), \qquad
\theta_{\hat{i}\hat{j}}= 0
\end{equation}
or
\begin{equation} \label{tnut-2nd-f-form}
a^{\dot{1}} = \frac{r_0}{r_0 + m} \sigma^1, \qquad  a^{\dot{2}} =
\frac{r_0}{r_0 + m} \sigma^2, \qquad a^{\dot{3}} = \frac{2m^2}{(r_0 + m)^2}
\sigma^3.
\end{equation}

Using Eqs. \eq{taub-nut-su2} and \eq{tnut-2nd-f-form}, we get the following result
\begin{eqnarray}
&& F^a \wedge F^a = 24 m^3 \frac{r-m}{(r+m)^5} \sigma^1
\wedge \sigma^2 \wedge \sigma^3  \wedge dr, \nonumber \\
&& a^a \wedge F^a|_{r=r_0} = -4 m^2 \frac{(r_0-m)^2}{(r_0+m)^4} \sigma^1
\wedge \sigma^2 \wedge \sigma^3, \nonumber \\
&& a^{\dot{1}} \wedge a^{\dot{2}} \wedge a^{\dot{3}} =
\frac{2 m^2 r_0^2}{(r_0+m)^4} \sigma^1
\wedge \sigma^2 \wedge \sigma^3.
\end{eqnarray}
Therefore, we see that the boundary integrals vanish because
\begin{equation}
a^a \wedge F^a |_{r_0 \to \infty} = 0, \qquad
a^{\dot{1}} \wedge a^{\dot{2}} \wedge a^{\dot{3}} |_{r_0 \to \infty} = 0.
\end{equation}
Finally we get the topological numbers for the Taub-NUT space
\begin{eqnarray} \la{nut-euler}
\chi(M) &=& \frac{1}{4\pi^2} \int_M F^a \wedge F^a \nonumber \\
&=& \frac{24 m^3}{4\pi^2} \underbrace{ \int_{{\bf S}^3} \sigma^1 \wedge  \sigma^2 \wedge
\sigma^3 }_{ = 16 \pi^2} \underbrace{ \int_m^\infty  \frac{r-m}{(r+m)^5}
dr }_{= \frac{1}{96 m^3}} = 1, \\
\la{nut-sign}
\tau(M) &=& \frac{1}{6\pi^2} \int_M F^a \wedge F^a  + \eta_S(\partial M) \nonumber \\
&=& \frac{2}{3}  + \eta_S(\partial M) = 0.
\end{eqnarray}
We have used the result (\ref{index-eta}) for the $\eta$-invariant with $k=1$.
In this case, the Euler number \eq{nut-euler} is equal to the instanton
number \eq{top-num-gi} because there is no boundary correction \ct{poyu,kim-yoon1}.
And it is straightforward to check the relation \eq{conjecture}.

\subsection{Eguchi-Hanson space}

For the product metric
\begin{equation}
ds^2 = \Big(1 - \frac{a^4}{r_0^4} \Big)^{-1} dr^2 + \frac{r_0^2}{4}
(\sigma_{1}^2 + \sigma_2^2) + \frac{r_0^2}{4} \Big(1 - \frac{a^4}{r_0^4} \Big) \sigma_3^2,
\end{equation}
the second fundamental form at the boundary $r = r_0$ is
\begin{equation} \label{eh-2nd-f-form}
a^{\dot{1}} = \frac{1}{2} \sqrt{1 - \frac{a^4}{r_0^4}} \sigma^1, \qquad  a^{\dot{2}} =
\frac{1}{2} \sqrt{1 - \frac{a^4}{r_0^4}} \sigma^2, \qquad a^{\dot{3}} =
\frac{1}{2} \Big(1 + \frac{a^4}{r_0^4} \Big) \sigma^3.
\end{equation}
Note that we have to choose the angular coordinate ranges
\begin{equation}
0 \leq \theta < \pi, \qquad 0 \leq \varphi < 2\pi, \qquad 0 \leq
\psi < 2\pi
\end{equation}
to remove the apparent singularities in the metric at $r=a$. Thus
the boundary at $\infty$ becomes $\mathbb{R}P^3$.

Then we obtain the following result
\begin{eqnarray}
&& F^a \wedge F^a = \frac{6a^8}{r^9} \sigma^1
\wedge \sigma^2 \wedge \sigma^3 \wedge dr, \nonumber \\
&& a^a \wedge F^a |_{r_0 \to \infty} = 0, \nonumber \\
&& a^{\dot{1}} \wedge a^{\dot{2}} \wedge a^{\dot{3}} |_{r_0 \to \infty} =
\frac{1}{8} \sigma^1 \wedge \sigma^2 \wedge \sigma^3,
\end{eqnarray}
and get the topological numbers for the Eguchi-Hanson space
\begin{eqnarray} \la{eh-euler}
\chi(M) &=& \frac{1}{4\pi^2} \int_M F^a \wedge F^a  + \frac{1}{12\pi^2}
\int_{\partial M}  \varepsilon^{abc} a^{a} \wedge a^{b}
\wedge a^{c} \nonumber \\
&=& \frac{6 a^8}{4 \pi^2} \underbrace{ \int_{{\bf RP}^3} \sigma^1 \wedge  \sigma^2 \wedge
\sigma^3 }_{ = 8 \pi^2} \underbrace{ \int_a^\infty  \frac{1}{r^9}
dr}_{= \frac{1}{8 a^8}} + \frac{6}{96\pi^2} \underbrace{ \int_{{\bf RP}^3}
\sigma^1 \wedge  \sigma^2 \wedge \sigma^3 }_{= 8 \pi^2} \nonumber \\
&=& \frac{3}{2} + \frac{1}{2} = 2, \\
\la{eh-sign}
\tau(M) &=& \frac{1}{6\pi^2} \int_M F^a \wedge F^a
+ \eta_S(\partial M) = 1.
\end{eqnarray}

Unlike the Taub-NUT case, there is a nontrivial boundary correction for the Euler
number \eq{eh-euler}. Since the instanton number \eq{top-num-gi} does not take
the boundary contribution into account,
it gets a fractional number $k = \frac{3}{2}$ \ct{bcc2,kim-yoon1}.
One can check that the relation \eq{conjecture} is satisfied.

\subsection{Euclidean Schwarzschild solution}

This solution is interesting because it has a nontrivial Euler number \ct{gi-hawking}
although it is not a gravitational instanton. But it turns out that this solution
is actually the sum of $SU(2)_L$ instanton and $SU(2)_R$ anti-instanton,
which explains why it has a nontrivial Euler number.

Take the product metric
\begin{equation}\label{prod-sbh}
    ds^2 = \Big(1 - \frac{2m}{r_0} \Big) d\tau^{2} + \Big(1 - \frac{2m}{r_0} \Big)^{-1} dr^{2}
+ r_0^{2} (d \theta^{2} + \sin \theta^2 d\phi^2).
\end{equation}
The second fundamental form at the boundary $r = r_0$ is then given by
\begin{equation} \label{sbh-2nd-f-form}
a^{\dot{1}} = - \frac{m}{r_0^2}d\tau, \qquad  a^{\dot{2}} = a^{\dot{3}} = 0.
\end{equation}
Using the result \eq{curv-esbh} with the definition $F^{(\pm)a} = \frac{1}{4}
\eta^{(\pm)a}_{AB} R_{AB}$, we obtain
\begin{eqnarray}
&& F^{(\pm)a} \wedge F^{(\pm)a} = \pm \frac{3m^2}{r^4} dr
\wedge d\Omega \wedge d\tau, \nonumber \\
&& a^{(\pm)a} \wedge F^{(\pm)a} |_{r_0 \to \infty} = 0, \nonumber \\
&& a^{(\pm)\dot{1}} \wedge a^{(\pm)\dot{2}} \wedge a^{(\pm)\dot{3}} = 0.
\end{eqnarray}

It is then straightforward to get the topological invariants \ct{gi-hawking}
\begin{eqnarray} \la{esbh-euler}
\chi(M) &=& \chi^+ (M) + \chi^- (M) = 2, \\
\la{esbh-sign}
\tau (M) &=& \tau^+ (M) - \tau^- (M) = 0,
\end{eqnarray}
where $\chi^+ (M) = \chi^- (M) = 1$ and $\tau^+ (M) = \tau^- (M)
= \frac{2}{3} + \eta(\partial M)$.
Hence we confirm that the Euclidean Schwarzschild solution \eq{e-sbh}
is the sum of an $SU(2)$ instanton and an anti-instanton.
And the relation \eq{top-rel-nc} implies that $\tau^\pm(M) = 0$ or
$\eta(\partial M) = - \frac{2}{3}$. Therefore the $SU(2)$ instanton for the Euclidean
Schwarzschild solution \eq{e-sbh} has the same topological invariants
as the Taub-NUT space \eq{taub-nut} \ct{duff}.
Note that two instantons belong to different gauge groups,
one in $SU(2)_L$ and the other in $SU(2)_R$,
and so they cannot decay into a vacuum.
As a result, the space \eq{e-sbh} should be stable at least perturbatively.
One may ask whether this kind of feature is special or general.
Remarkably it can be shown \cite{our-next} that any Ricci-flat four-manifold always arises as the sum of $SU(2)_L$ instantons and $SU(2)_R$ anti-instantons.
Hence any Ricci-flat manifold should be stable
for the same reason.

\subsection{Topological invariant of Yang-Mills instantons}

We have noticed that the instanton number \eq{top-num-su2} for (anti-)self-dual
gauge fields satisfying \eq{su2-instanton} is not necessarily integer-valued
because it does not take possible boundary corrections into account.
But the equivalence of the self-dual systems in \eq{su2-instanton} and
\eq{g-instanton} implies that we need to also consider boundary contributions
for the topological charge of Yang-Mills instantons defined on a curved manifold.
Thereby we suggest the Chern number for an instanton bundle
including boundary corrections
\begin{equation} \label{instanton-number}
k = \frac{1}{16 \pi^2} \int_M  F^{a}  \wedge F^{a}
+ \frac{1}{16 \pi^2} \int_{\partial M_\infty} A^{a} \wedge F^{a}
- \frac{1}{96 \pi^2} \int_{\partial M_\infty}
\varepsilon^{abc} A^{a} \wedge A^{b} \wedge A^{c}
\end{equation}
which can be identified with the Euler characteristic \eq{euler-gauge} in the self-dual gauge, $A^{(-)a}=0$, with the gauge theory normalization $A_{YM}^a = - 2 A_{G}^a$
and $F_{YM}^a = - 2 F_{G}^a$ and is accordingly integer-valued.
Note that the boundary term in \eq{instanton-number} is precisely the Chern-Simons form 
for the $SU(2)$ vector bundle at an asymptotic infinity.

Now we consider the four-manifold $M$ to have two ends,
one at an asymptotic infinity $\partial M_\infty$ and
the other at an inner boundary $\partial M_0$ describing nuts and bolts of
gravitational instantons \ct{gh-cmp}.
For example, the inner boundary is at $r=m$ for the Taub-NUT space \eq{taub-nut}
and at $r=a$ for the Eguchi-Hanson space \eq{eguchi-hanson}.
Using the identity $F^{a}  \wedge F^{a} = dK$ where
\begin{equation}\label{chern-simons}
    K =  A^{a} \wedge dA^{a} + \frac{1}{3} \varepsilon^{abc} A^{a}
    \wedge A^{b} \wedge A^{c}
\end{equation}
and the boundary operation $\partial M = \partial M_0 - \partial M_\infty$,\footnote{The sign is due to our choice of orientation. See the footnote \ref{orientation}.}
one can rewrite the instanton number \eq{instanton-number} as the Chern-Simons integral
on the inner boundary $\partial M_0$, i.e.,
\begin{equation} \label{cs-instanton-number}
k = \frac{1}{16 \pi^2} \int_{\partial M_0}
\Big(A^{a} \wedge F^{a} - \frac{1}{6} \varepsilon^{abc} A^{a}
\wedge A^{b} \wedge A^{c} \Big).
\end{equation}
Recall that the instanton number \eq{instanton-number} is simply the expression of
the Euler number \eq{euler-gauge} and the Euler number $\chi(M)$ can be determined
by the set of nuts and bolts through the fixed point theorem (Eq. (4.6) in \ct{gh-cmp})
\begin{equation}\label{euler-bn}
   \chi(M) = \sharp (\rm{nuts}) + 2 \; \sharp (\rm{bolts}).
\end{equation}
Then we get a very interesting result that the Chern-Simons integral \eq{cs-instanton-number} on the inner boundary $\partial M_0$ simply counts
the number of nuts plus the twice of the number of bolts in gravitational instantons:
\begin{equation} \label{cs-nut-bolt}
k = \frac{1}{16 \pi^2} \int_{\partial M_0}
\Big(A^{a} \wedge F^{a} - \frac{1}{6} \varepsilon^{abc} A^{a}
\wedge A^{b} \wedge A^{c} \Big) = \sharp (\rm{nuts}) + 2 \; \sharp (\rm{bolts}).
\end{equation}
It is easy to check the result \eq{cs-nut-bolt} for the Taub-NUT space
$(\sharp (\rm{nuts})=1, \; \sharp (\rm{bolts}) = 0)$ and
for the Eguchi-Hanson space $(\sharp (\rm{nuts})=0, \; \sharp (\rm{bolts}) = 1)$,
using the previous results with the relation $A_{YM}^a = - 2 A_{G}^a$
and $F_{YM}^a = - 2 F_{G}^a$.

\section{Discussion}

Let us go back to the questions we have raised in Section 1.
So far we have focused on the similarity between gauge theory and gravitation.
A main source of the similarity is coming from the fact that the $O(4)$-valued 1-forms
${\omega^A}_B$ are gauge fields (a connection of the spin bundle $SM$) with respect to $O(4)$ rotations as shown in Eq. \eq{spin-so4}.
Then the Riemann curvature tensors in \eq{cartan-curvature}
constitute $O(4)$-valued curvature 2-forms of the spin bundle $SM$.
Therefore, the four-dimensional Euclidean gravity can be formulated as a gauge theory
using the language of the $O(4)$ gauge theory.
Via the fact that the Lorentz group $O(4)$ is a direct product of normal subgroups
$SU(2)_L$ and $SU(2)_R$, i.e. $O(4)= SU(2)_L \times SU(2)_R$,
the four-dimensional Euclidean gravity can be decomposed into two copies of $SU(2)$ gauge theories. In particular, the (anti-)self-dual sector satisfying \eq{g-instanton} can be
formulated as an $SU(2)$ gauge theory, as clearly indicated in Eq. \eq{spin-su2}.

Nevertheless, gravity is different from gauge theory in many aspects.
A decisive source of the difference is the existence of a Riemannian metric
which does not have any counterpart in gauge theory.
We highlight some crucial differences
between gauge theory and gravitation with the following table:
\begin{center}
\begin{tabular}{|c|c|c|}
  \hline
   Property   & Einstein  & Yang-Mills \\
  \hline
  Metric & $g_{MN} (x)$ or $E^A$ & $\ldots$ \\
  \hline
  Torsion & $dE^A + {\omega^A}_B \wedge E^B = 0 $  & $\ldots$ \\
  \hline
  Cyclic identity & ${R^A}_B \wedge E^B = 0 $ & $\ldots$  \\
  \hline
  Einstein equation  & $G_{MN} = 8 \pi G T_{MN}$  & $\ldots$ \\
  \hline
  Coupling constant  & $[G]= L^2$ & $[g_{YM}] = L^0$ \\
  \hline
  Symmetry  & Spacetime  & Internal \\
  \hline
  Interaction & Long-range & Short-range \\
  \hline
\end{tabular}
\end{center}
The metric is constrained to be covariantly constant with respect to the Levi-Civita connection \eq{levi-civita} or equivalently the vierbeins are constrained
to be torsion-free, i.e.,  $T^A=dE^A + {\omega^A}_B \wedge E^B = 0$.
This constraint leads to the result that the spin connections ${\omega^A}_B$
are determined by potential fields, i.e., vierbeins, as Eq. \eq{spin-conn}.
As a result, a primary field for gravity is the metric tensor
rather than a gauge field (a connection of vector bundle).
This extra structure comprises a core origin of the differences in the above table.

Recently one of us showed \ct{hsy-jhep} (see also recent reviews \cite{jjl-hsy} and \cite{hsy-review}) that Einstein gravity can be derived from electromagnetism in noncommutative space. In particular, the vierbeins $E_A$ in gravity arise
from the leading order of noncommutative $U(1)$ gauge fields
and higher order terms give rise to derivative corrections to Einstein gravity.
Actually the Einstein equations arising from the noncommutative gauge fields
and the resulting emergent gravity motivate to newly address
the questions in Section 1 in a more broad context to include noncommutative $U(1)$
gauge theories. For example, it was rigorously shown in \ct{hsy-instanton} that noncommutative $U(1)$ instantons are equivalent to gravitational instantons.
Therefore, it will be very interesting to find a precise map between noncommutative $U(1)$ instantons and Yang-Mills instantons because a particular class of Yang-Mills instantons
can be obtained from gravitational instantons as was shown in this paper.
We hope to draw some valuable insights from this line of thought in our future works.

Now our method in Section 3 can easily be generalized to get new instanton solutions
by the conformal rescaling method \ct{ete-hau}. Suppose that $(M, g)$ is a self-dual gravitational instanton and consider a Weyl transformation given by Eq. \eq{conf-tr}
which can be represented as $\widetilde{E}^A = \Omega(x) E^A \in \Gamma(T^*M)$
or $\widetilde{E}_A = \Omega^{-1}(x) E_A \in \Gamma(TM)$ in terms of vierbeins.
Under the Weyl transformation, the spin connections transform as follow:
\begin{equation}\label{spin-weyl}
\widetilde{\omega}_{AB} = \omega_{AB} + (E_B \log \Omega E^A - E_A \log \Omega E^B).
\end{equation}
We can apply the decompositions \eq{spin-sd-asd} and
\eq{curvature-sd-asd} to the transformed spin connection \eq{spin-weyl} and the corresponding curvature tensor $\widetilde{R} = d \widetilde{\omega} + \widetilde{\omega} \wedge \widetilde{\omega}$, respectively. After all, we will get new $SU(2)$ gauge fields
defined by
\begin{equation}\label{gauge-weyl}
    \widetilde{A}^{(+)a} = A^{(+)a} + \mathfrak{A}^{(+)a}, \qquad
    \widetilde{A}^{(-)a} = \mathfrak{A}^{(-)a}
\end{equation}
where $A^{(+)a}$ are the self-dual gauge fields determined by the original
self-dual spin connection $\omega_{AB}$ and
\begin{equation}\label{new-weyl}
    \mathfrak{A}^{(\pm)a} \equiv \frac{1}{2} \eta^{(\pm)a}_{AB}
    (E_B \log \Omega) E^A
\end{equation}
and the corresponding $SU(2)$ field strengths will be given by
\begin{equation}\label{new-curvature}
    \widetilde{F}^{(\pm)a} = d \widetilde{A}^{(\pm)a} - \varepsilon^{abc} \widetilde{A}^{(\pm)b} \wedge \widetilde{A}^{(\pm)c}.
\end{equation}

Now we can make two different choices:\footnote{It may be worthwhile to compare
the solution \eq{new-weyl} with 't Hooft ansatz (see Sect. 4.3. in \ct{rajaraman})
in singular (the case (I)) and regular (the case (II)) gauges. Note that the solution \eq{su2-gauge} from the Gibbons-Hawking metric takes the form \eq{new-weyl} for the case (I).}
\begin{eqnarray} \la{weyl-case1}
&& {\rm (I)} \; \widetilde{F}^{(-)a}_{AB} = \frac{1}{2}{\varepsilon_{AB}}^{CD} \widetilde{F}^{(-)a}_{CD} , \\
\la{weyl-case2}
&& {\rm (II)} \; \widetilde{F}^{(-)a}_{AB} = - \frac{1}{2}{\varepsilon_{AB}}^{CD} \widetilde{F}^{(-)a}_{CD}.
\end{eqnarray}
For the first choice (I), we will get a self-dual Yang-Mills instanton
while, for the second choice (II), an anti-self-dual Yang-Mills instanton.
Then one can show \ct{our-next} that, for the case (I),
the Ricci-scalar $\widetilde{R} = \widetilde{g}^{MN} \widetilde{R}_{MN}$
will identically vanish, i.e. $\widetilde{R} = 0$, but the case (II) seems to give rise
to an intriguing manifold satisfying $\widetilde{R}_{MN} - \frac{1}{4}\widetilde{g}_{MN} \widetilde{R} = 0$. Because the Ricci scalar transforms under the Weyl transformation \eq{conf-tr} as $\Omega^3 \widetilde{R} = \Omega R - 6 \Box_g \Omega$
where $\Box_g$ refers to the scalar Laplacian on $(M, g)$,
we see that the rescaling function $\Omega(x)$ must be harmonic, i.e.
$\Omega^{-1} \Box_g \Omega = 0$, for the case (I), taking into account that $R=0$.
But the harmonic function $\Omega(x)$ will allow mild singularities \ct{ete-hau}
which can be removed by a gauge transformation.

By the same procedure as Eq. \eq{ym-instanton-cod}, the self-dualities in Eqs. \eq{weyl-case1} and \eq{weyl-case2} can be written as
\begin{equation} \label{tym-weyl}
\widetilde{F}^{(-)}_{MN} = \pm  \frac{1}{2}
\frac{\varepsilon^{RSPQ}}{\sqrt{\widetilde{g}}} \widetilde{g}_{MR} \widetilde{g}_{NS}
\widetilde{F}^{(-)}_{PQ}
\end{equation}
where $\sqrt{\widetilde{g}} = \Omega^4 \sqrt{g}$.
However, taking into account the conformal invariance of self-duality,
we get the self-duality equation on the original four-manifold $(M,g)$, i.e.,
\begin{equation} \label{ym-weyl}
\widetilde{F}^{(-)}_{MN} = \pm  \frac{1}{2} \frac{\varepsilon^{RSPQ}}{\sqrt{g}} g_{MR} g_{NS} \widetilde{F}^{(-)}_{PQ}.
\end{equation}
Consequently, we get new Yang-Mills instantons on an original Ricci-flat manifold
$(M,g)$ after the Weyl transformation \eq{spin-weyl}.
More details about explicit solutions obtained in this way and
their topological properties will be discussed elsewhere.

In this paper we showed that any gravitational instanton is an $SU(2)$ Yang-Mills
instanton on the gravitational instanton itself. Regarding to this property,
there is an interesting theorem (Example 3 (page 302) in \ct{kron-naka} and see also
Sect. 7 in \ct{italy}) that there always exists an instanton bundle on an ALE manifold $M$
with the instanton number $k = 1 - \frac{1}{|\Gamma|}$ ($|\Gamma|$ denoting the order
of $\Gamma$ in $M \cong \widetilde{\mathbb{C}^2/\Gamma}$) defined by \eq{top-num-su2}
such that the moduli space of self-dual connections on the instanton bundle is a
four-dimensional hyper-K\"ahler manifold and coincides with the base manifold $M$.
Inferred from our result, the above property seems to be true
for other self-dual manifolds.
To be precise, suppose that $\mathcal{M}(E \to M, k)$ is the moduli space of self-dual
connections on a vector bundle $E$ over $M$ with instanton number $k$
where $M$ is a gravitational instanton.
Then, each non-empty, non-compact 4-dimensional component of
the moduli space $\mathcal{M}(E \to M, k)$ is isomorphic to the gravitational
instanton itself. It will be interesting to clarify this assertion.

\section*{Acknowledgments}

We are grateful to Kimyeong Lee for guiding us to a correct direction.
HSY thanks Jungjai Lee for helpful discussions for some issues.
The work of C. Park was supported by the National Research Foundation of Korea (NRF) grant
funded by the Korea government (MEST) through the Center for Quantum Spacetime (CQUeST)
of Sogang University with grant number 2005-0049409.
The work of H.S. Yang was supported by the RP-Grant 2009 of Ewha Womans University.

%\newpage

%%%%%%%%%%%%%%%%% Journal Macros JHEP Style %%%%%%%%%%%%%%%%%%%%%%%%%%%
\nc{\PR}[3]{Phys. Rev. {\bf #1}, #2 (#3)}
\nc{\NPB}[3]{Nucl. Phys. {\bf B#1}, #2 (#3)}
\nc{\PLB}[3]{Phys. Lett. {\bf B#1}, #2 (#3)}
\nc{\PRD}[3]{Phys. Rev. {\bf D#1}, #2 (#3)}
\nc{\PRL}[3]{Phys. Rev. Lett. {\bf #1}, #2 (#3)}
\nc{\PREP}[3]{Phys. Rep. {\bf #1}, #2 (#3)}
\nc{\EPJ}[3]{Eur. Phys. J. {\bf C#1}, #2 (#3)}
\nc{\PTP}[3]{Prog. Theor. Phys. {\bf #1}, #2 (#3)}
\nc{\CMP}[3]{Commun. Math. Phys. {\bf #1}, #2 (#3)}
\nc{\MPLA}[3]{Mod. Phys. Lett. {\bf A#1}, #2 (#3)}
\nc{\CQG}[3]{Class. Quant. Grav. {\bf #1}, #2 (#3)}
\nc{\NCB}[3]{Nuovo Cimento {\bf B#1}, #2 (#3)}
\nc{\ANNP}[3]{Ann. Phys. (N.Y.) {\bf #1}, #2 (#3)}
\nc{\GRG}[3]{Gen. Rel. Grav. {\bf #1}, #2 (#3)}
\nc{\MNRAS}[3]{Mon. Not. Roy. Astron. Soc. {\bf #1}, #2 (#3)}
\nc{\JHEP}[3]{J. High Energy Phys. {\bf #1}, #2 (#3)}
\nc{\JCAP}[3]{JCAP {\bf #1}, #2 {#3}}
\nc{\ATMP}[3]{Adv. Theor. Math. Phys. {\bf #1}, #2 (#3)}
\nc{\AJP}[3]{Am. J. Phys. {\bf #1}, #2 (#3)}
\nc{\ibid}[3]{{\it ibid.} {\bf #1}, #2 (#3)}
\nc{\ZP}[3]{Z. Physik {\bf #1}, #2 (#3)}
\nc{\PRSL}[3]{Proc. Roy. Soc. Lond. {\bf A#1}, #2 (#3)}
\nc{\LMP}[3]{Lett. Math. Phys. {\bf #1}, #2 (#3)}
\nc{\AM}[3]{Ann. Math. {\bf #1}, #2 (#3)}
\nc{\hepth}[1]{{\tt [arXiv:hep-th/{#1}]}}
\nc{\grqc}[1]{{\tt [arXiv:gr-qc/{#1}]}}
\nc{\astro}[1]{{\tt [arXiv:astro-ph/{#1}]}}
\nc{\hepph}[1]{{\tt [arXiv:hep-ph/{#1}]}}
\nc{\phys}[1]{{\tt [arXiv:physics/{#1}]}}
\nc{\arxiv}[1]{{\tt [arXiv:{#1}]}}

%%%%%%%%%%%%%%%%%%%%%%%%%%%%%%%%%%%%%%%%%%%%%%%%%%%%%%%%%%%


\begin{thebibliography}{99}
%%%%%%%%% References %%%%%%%%%%%%%%%%%%%%%%%%%%%%%%%%%%%%%%%%%%%%%%


\bibitem{rajaraman} R. Rajaraman, {\it Solitons and Instantons}
(North Holland, Amsterdam, 1982).



\bibitem{swn=2} N. Seiberg and E. Witten, Nucl. Phys. {\bf B426}, 19 (1994);
Nucl. Phys. {\bf B431}, 484 (1994).



\bibitem{adhm} M. F. Atiyah, N. J. Hitchin, V. G. Drinfeld and Yu. I. Manin,
Phys. Lett. {\bf 65A}, 185 (1978).



\bibitem{inst-review1} N. Dorey, T. J. Hollowood, V. V. Khoze and M. P. Mattis
Phys. Rept. {\bf 371}, 231 (2002).



\bibitem{inst-review2} E. J. Weinberg and P. Yi, Phys. Rept. {\bf 438}, 65 (2007).




\bibitem{nekrasov} N. A. Nekrasov, Adv. Theor. Math. Phys. {\bf 7}, 831 (2004).




\bibitem{nek-oko} N. A. Nekrasov and A. Okounkov,
Seiberg-Witten Theory and Random Partitions, \hepth{0306238}.




\bibitem{donaldson} S. K. Donaldson and P. B. Kronheimer,
{\it The Geometry of Four-Manifolds} (Oxford University Press, 1990).




\bibitem{gi-hawking} S. W. Hawking, Phys. Lett. {\bf 60A}, 81 (1977).




\bibitem{egh-report} T. Eguchi, P. B. Gilkey and A. J. Hanson,
Phys. Rep. {\bf 66}, 213 (1980).




\bibitem{gibbs-pope} G. W. Gibbons and C. N. Pope, Commun.
Math. Phys. {\bf 66}, 267 (1979).




\bibitem{big-book} C. W. Misner, K. S. Thorne and J. A. Wheeler,
{\it Gravitation} (W. H. Freeman and Company, New York, 1973).




\bibitem{eh-ap} T. Eguchi and A. J. Hanson, Annals Phys. {\bf 120}, 82 (1979).




\bibitem{hitchin} N. J. Hitchin, Math. Proc. Camb. Phil. Soc. {\bf 85}, 465 (1979).




\bibitem{kronheimer} P. B. Kronheimer, J. Diff. Geom. {\bf 29}, 665 (1989).




\bibitem{gibbons} G. W. Gibbons, P. Rychenkova and R. Goto,
Commun. Math. Phys. {\bf 186}, 585 (1997).




\bibitem{hitchin-cherkis} S. A. Cherkis and N. J. Hitchin,
Commun. Math. Phys. {\bf 260}, 299 (2005).





\bibitem{kummer-gi} O. Biquard and V. Minerbe, A K\"ummer construction
for gravitational instantons, \arxiv{1005.5133}.




\bibitem{kron-naka} P. B. Kronheimer and H. Nakajima, Ann. Math. {\bf 288}, 263 (1990).




\bibitem{italy} M. Bianchi, F. Fucito, G. Rossi and M. Martellini,
Nucl. Phys. {\bf B473}, 367 (1996).




\bibitem{cherkis-alf} S. A. Cherkis, Commun. Math. Phys. {\bf 290}, 719 (2009);
Adv. Theor. Math. Phys. {\bf 14}, 609 (2010).




\bibitem{witten-alf} E. Witten, J. High Energy Phys. {\bf 06}, 067 (2009).




\bibitem{sw-strong} C. Vafa and E. Witten, \NPB{431}{3}{1994}.




\bibitem{ale-instanton} M. R. Douglas and G. Moore, D-branes, Quivers, and ALE Instantons,
\hepth{9603167}.




\bibitem{hsy-jhep} H. S. Yang, \JHEP{05}{012}{2009}.




\bibitem{jjl-hsy} J. Lee and H. S. Yang,  Quantum Gravity from Noncommutative Spacetime,
\arxiv{1004.0745}.




\bibitem{duff} J. M. Charap and M. J. Duff, Phys. Lett. {\bf 69B}, 445 (1977);
Phys. Lett. {\bf 71B}, 219 (1977).




\bibitem{het-inst} M. Bianchi, F. Fucito, G. C. Rossi and M. Martellini,
Nucl. Phys. {\bf B440}, 129 (1995).




\bibitem{bcc1} H. Boutaleb-Joutei, A. Chakrabarti and A. Comtet,
Phys. Rev. {\bf D21}, 979 (1980).




\bibitem{bcc2} H. Boutaleb-Joutei, A. Chakrabarti and A. Comtet,
Phys. Rev. {\bf D21}, 2280 (1980).




\bibitem{poyu} C. N. Pope and A. L. Yuille, Phys. Lett. {\bf 78B}, 424 (1979).




\bibitem{kim-yoon1} H. Kim and Y. Yoon, Phys. Lett. {\bf B495}, 169
(2000).




\bibitem{gh-cmp} G. W. Gibbons and S. W. Hawking, \CMP{66}{291}{1979}.




\bibitem{gpr-index} G. W. Gibbons, C. N. Pope and H. R\"omer, Nucl. Phys. {\bf B157},
377 (1979).




\bibitem{gibb-perry} G. W. Gibbons and M. J. Perry, Phys. Rev. {\bf D22}, 313 (1980).




\bibitem{buchdahl} N. P. Buchdahl, J. Diff. Geom. {\bf 24}, 19  (1986).




\bibitem{i-witten} E. Witten, Phys. Rev. Lett. {\bf 38}, 121 (1977).




\bibitem{ete-hau} G. Etesi and T. Hausel, Phys. Lett. {\bf B514}, 189
(2001); Commun. Math. Phys. {\bf 235}, 275 (2003).





\bibitem{kim-yoon2} H. Kim and Y. Yoon, Phys. Rev. {\bf D63}, 125002
(2001); Phys. Rev. {\bf D63}, 126003 (2001).




\bibitem{radu} Y. Brihaye and E. Radu, Europhys. Lett. {\bf 75}, 730 (2006);
E. Radu, D. H. Tchrakian and Y. Yang, Phys. Rev. {\bf D77}, 044017 (2008);
O. Mi\v skovi\'c and R. Olea, Phys. Rev. {\bf D79}, 124020 (2009).




\bibitem{g-review} M. S. Volkov and D. V. Gal'tsov, Phys. Rept. {\bf 319}, 1 (1999);
S. A. Cherkis, Instantons on Gravitons, \arxiv{1007.0044}.




\bibitem{dunajski} M. Dunajski, {\it Solitons, Instantons and Twistors}
(Oxford University Press, Oxford, 2010).





\bibitem{taubes} C. H. Taubes, J. Diff. Geom. {\bf 19}, 517 (1984).




\bibitem{tsukamoto} M. Tsukamoto, An open four-manifold having no instanton,
\arxiv{1004.3394}.




\bibitem{gh-instanton} G. W. Gibbons and S. W. Hawking, Phys. Lett. {\bf 78B}, 430 (1978).




\bibitem{eh-inst} T. Eguchi and A. J. Hanson, Phys. Lett. {\bf 74B}, 249 (1978).




\bibitem{ah-instanton} M. F. Atiyah and N. J. Hitchin, Phys. Lett. {\bf 107A}, 21 (1985).




\bibitem{hana-piol} A. Hanany and B. Pioline, J. High Energy Phys. {\bf 07}, 001 (2000).




\bibitem{real-heaven} C. P. Boyer and J. D. Finley, J. Math. Phys. {\bf 23}, 1126 (1982).




\bibitem{our-next} J. J. Oh and H. S. Yang, Einstein Manifolds As Yang-Mills Instantons,
\arxiv{1101.5185}.




\bibitem{italy2} D. Anselmi, M. Bill\'o, P. Fr\'e, L. Girardello and
A. Zaffaroni, Int. J. Mod. Phys. {\bf A9}, 3007 (1994).




\bibitem{hsy-review} H. S. Yang, Mod. Phys. Lett. {\bf A25}, 2381 (2010).




\bibitem{hsy-instanton} M. Salizzoni, A. Torrielli and H. S. Yang,
Phys. Lett. {\bf B634}, 427 (2006) 427;
H. S. Yang and M. Salizzoni, Phys. Rev. Lett. {\bf 96}, 201602 (2006);
H. S. Yang, Europhys. Lett. {\bf 88}, 31002 (2009);
Eur. Phys. J. {\bf C64}, 445 (2009).



\end{thebibliography}
\end{document}